\documentclass[pre,twocolumn,showpacs,superscriptaddress]{revtex4-2}
\usepackage{amsfonts,amssymb,amsmath}
\usepackage{graphicx}
\usepackage{bm}
\usepackage{xcolor}
\usepackage{mathrsfs}
\usepackage{hyperref}
\usepackage{gensymb}
\usepackage{booktabs}
\usepackage{amsmath,amssymb,bm}
\usepackage[ruled,vlined]{algorithm2e}
\SetKw{Break}{break}
\newcommand{\vect}[1]{\bm{#1}}
\newcommand{\lap}{\mathscr{L}} % Laplacian matrix
\newcommand{\matD}{\mathbf{D}} % Degree matrix
\newcommand{\matA}{\mathbf{A}} % Adjacency matrix
\newcommand{\matI}{\mathbf{I}}
\newcommand{\matP}{\mathbf{P}}
\usepackage{float}

\renewcommand{\today}{%
  \number\day\space%
  \ifcase\month\or
    January\or February\or March\or April\or May\or June\or
    July\or August\or September\or October\or November\or December\fi
  \space \number\year
}

\begin{document}

\title{
Spectral Gradient Iterative Edge Attack for Synchronization Suppression in Complex Networks
}
\author{Kaiming Luo}
\thanks{Corresponding author: kmluo24@m.fudan.edu.cn}
\affiliation{School of Information Science and Technology, Fudan University, Shanghai, 200438, China}

\date{Received: date  \today}

\begin{abstract}
The synchronization of complex networks, governed by the generalized Fiedler value ($\gamma$) of the Laplacian matrix, is critical for functional stability and energy efficiency. However, this property also renders networks vulnerable to targeted disruptions. Traditional percolation-based attack strategies, which focus on structural integrity, often fail to effectively suppress synchronization. This study introduces a Laplacian spectral perturbation approach to systematically identify and remove edges critical to synchronization. By deriving the sensitivity of $\gamma$ to topological changes and leveraging the gradient of the Fiedler vector, we quantify each edge's contribution to synchronization, revealing its connection to community structure. We propose the \emph{Fiedler Gradient Iterative Attack (FGIA)} algorithm for static networks, which constructs locally optimal edge removal sequences to maximize $\gamma$ degradation while preserving global connectivity. FGIA achieves computational efficiency, outperforming brute-force methods and conventional centrality-based attacks. Extensive simulations on synthetic and real-world networks demonstrate FGIA's superior performance in synchronization suppression, offering practical applications in neuroscience and critical infrastructure protection.

\end{abstract}

\maketitle
% \noindent{\it Keywords}: synchronization dismantling, algebraic connectivity, spectral perturbation  theory, edge removal algorithm

\textbf{Synchronization is a hallmark of collective behavior in complex systems, governing phenomena from neural coherence to power-grid stability. Yet the same synchrony that enables efficient coordination can also magnify systemic fragility, making its controlled suppression a fundamental problem across disciplines. This study develops a spectral-gradient framework that quantifies how each network edge contributes to synchronization stability via the Fiedler eigenmode of the Laplacian. Building on this insight, we propose the Fiedler Gradient Iterative Attack (FGIA) algorithm, which efficiently identifies and removes the most synchronization-critical edges while preserving overall connectivity. Extensive tests on synthetic and real-world networks demonstrate that FGIA surpasses conventional centrality-based and percolation-based attacks, providing a general, physically interpretable tool for analyzing and mitigating synchronization vulnerabilities in complex dynamical systems.}

\section{Introduction}

Synchronization is a fundamental manifestation of collective behavior in complex systems, playing a central role in ensuring functional robustness, rapid recovery, and adaptive reconfiguration following external disturbances~\cite{Arenas2008,Boccaletti2006PhysRep,Barahona2002,Qi2024NetworkResilience,Motter2005PRE}. 
Across a wide range of application domains, coherent dynamical coordination underpins mission-critical operations and system-level efficiency. 
At the same time, synchronization exhibits an intrinsic duality: while moderate synchrony enhances stability and performance, excessive or overly rigid synchronization can amplify systemic fragility, facilitating the propagation of local perturbations into large-scale cascading failures~\cite{Motter2015}. 
This ambivalent nature of synchronization has been documented in diverse contexts, including neuronal and brain networks~\cite{Belykh2005synchronization,Galinsky2023critically,mathiopoulou2025gamma}, infrastructure and power systems~\cite{ruan2023super,mishra2023resilience,Tsang2019}, and distributed artificial intelligence systems~\cite{Richard202Synchronization,chen2015saturated,pan2020justinian}, highlighting its broad interdisciplinary relevance.

A crucial insight emerging from prior studies is that structural robustness and functional (synchronization) robustness are not necessarily aligned~\cite{Motter2015}. 
In many networked systems, global connectivity may remain intact even when the capacity for collective synchronization is substantially degraded~\cite{Barahona2002,Arenas2008}. 
This mismatch arises because synchronization stability is governed by global spectral properties—most notably the Laplacian eigenvalue gap—rather than by purely local or combinatorial structural measures~\cite{Arenas2006,Barahona2002}. 
As a result, identifying which specific edges or nodes are most critical to sustaining collective synchrony remains a nontrivial challenge. 
The problem is further exacerbated in real-world networks, where community organization, directionality, and heterogeneous coupling weights profoundly influence how structural perturbations translate into dynamical consequences~\cite{Arenas2008}. 
Despite these complexities, many existing attack and robustness assessment strategies continue to rely primarily on static structural indicators, leaving functional vulnerabilities in network synchronization insufficiently characterized~\cite{Crucitti2004}. Moreover, these strategies are often formulated for undirected and unweighted graphs, limiting their applicability to realistic networks with directional and weighted interactions. 

Heuristic attacks based on local centrality measures—such as degree or betweenness—can exhibit good performance in specific settings, but their effectiveness is highly topology-dependent. 
For example, degree-based removal is particularly effective in scale-free networks~\cite{Albert2000Attack}, yet becomes markedly suboptimal in small-world or random network structures. 
This discrepancy between structural robustness and dynamical resilience underscores the need for topology-agnostic methodologies capable of systematically suppressing synchronization across directed and weighted networks.

A deeper conceptual challenge lies in distinguishing structural collapse from functional desynchronization. 
Whereas the GCC provides a measure of topological integrity, functional synchronization is governed by spectral properties of the Laplacian matrix, in particular the algebraic connectivity or Fiedler value $\lambda_2$~\cite{fiedler1973algebraic,Dorfler2013Sync}. 
Extending spectral synchronization metrics to directed or weighted networks introduces additional difficulty, as the Laplacian becomes asymmetric and may exhibit complex-valued eigenvalues, complicating the interpretation of spectral gaps and convergence rates. 
Recent studies have explored $\lambda_2$-based indicators of functional criticality at the level of nodes~\cite{chen2024node}, edges~\cite{jiang2024fiedler}, and higher-order structures such as cycles~\cite{jiang2023searching}. 
However, these approaches are typically static, overlooking the nonlinear dependence of synchronization capacity on successive structural perturbations. 
Furthermore, optimization strategies based on exhaustive enumeration of $\lambda_2$ degradation are computationally prohibitive for large-scale networks and neglect inter-edge correlations that naturally arise during progressive attacks.

In this work, we develop a spectral perturbation framework that analytically characterizes how structural modifications alter network synchronizability, with particular emphasis on the evolution of the second-smallest Laplacian eigenvalue $\lambda_2$. 
We show that the gradient of the Fiedler eigenvector—quantifying the spatial variation of the slowest dynamical mode—provides a direct, physically interpretable measure of an edge’s contribution to collective synchronization. 
For undirected networks, this formulation naturally exploits eigenvector orthogonality to isolate individual spectral contributions. 
For directed networks, the analysis generalizes through the biorthogonality condition between left and right eigenvectors, $y_k^{\top}x_l=\delta_{kl}$, enabling a unified treatment of symmetric and asymmetric topologies.

Building on this theoretical foundation, we propose the Fiedler Gradient Iterative Attack (FGIA), a locally optimal edge-removal algorithm designed to progressively suppress network synchronization while explicitly preserving global connectivity. 
FGIA offers three key advantages: (i) it systematically identifies edges whose removal maximally degrades synchronization capacity; (ii) it enforces connectivity constraints throughout the attack process; and (iii) it achieves high computational efficiency, significantly outperforming exhaustive optimization schemes and enabling scalability to large networks.

In contrast to enumeration-based optimization approaches~\cite{ZhangZeng2014PhysicaA,Richter2023SpectralDynamics,MilaneseSunNishikawa2010PRE,ChenLuZhanChen2012Chapter} and purely numerical attack strategies~\cite{YangJiang2017PhysicaA,ZengLuZhou2012NJP}, FGIA provides a physically transparent and computationally tractable pathway to synchronization suppression. 
Extensive simulations on both synthetic benchmarks and real-world networks demonstrate that FGIA consistently outperforms conventional topology-based heuristics, achieving comparable or superior functional disruption with substantially fewer edge removals. 
These results establish FGIA as a scalable, spectrally grounded framework for analyzing and controlling synchronization resilience in complex networked systems.

\section{Synchronization metric}
An essential metric for evaluating network synchronization is the convergence rate~\cite{Chung1997SpectralGraphTheory, Mohar1991LaplacianSpectrum,jiang2024fiedler},  which characterizes the speed at which the system returns to its steady state after a perturbation. Consider a network consisting of \( n \) agents, each with a \( d \)-dimensional state vector. Let \( \bm{\psi}(t) = [\psi_1(t)^\top, \dots, \psi_n(t)^\top]^\top \in \mathbb{R}^{n \times d} \) denote the collective state of the system at time \( t \), where \( \psi_i(t) \in \mathbb{R}^d \) represents the state of agent \( i \). The dynamics of the coupled system are described by
\begin{equation}
    \dot{\bm{\psi}} = F(\bm{\psi}) - (\lap \otimes \mathbf{H}) \bm{\psi},
    \label{eq:dynamics}
\end{equation}
where \( F: \mathbb{R}^{n \times d} \to \mathbb{R}^{n \times d} \) captures the intrinsic nonlinear dynamics of individual agents, \( \mathbf{H} \in \mathbb{R}^{d \times d} \) is the inner coupling matrix that governs interactions between components of agent states, and \( \lap \) denotes the Laplacian matrix that reflects the underlying network topology. The Kronecker product \( \otimes \) ensures consistency in dimensions when coupling across both node and state-space levels.

The Laplacian matrix \( \lap \in \mathbb{R}^{n \times n} \) is defined as \( \lap := \matD - \matA\), where \( \matA = [a_{ij}] \) is the adjacency matrix encoding pairwise coupling strengths and \( \matD = \operatorname{diag}(d_1, \dots, d_n) \) is the degree matrix with \( d_i = \sum_{j=1}^n a_{ij} \). This formulation is general and applies to both weighted and unweighted networks, as well as to directed and undirected topologies.

In the case of unweighted networks, the adjacency matrix \(\mathbf{A}\) is binary, with \(a_{ij} = 1\) if there is a connection from node \(j\) to node \(i\), and \(a_{ij} = 0\) otherwise. For weighted networks, \(a_{ij} \geq 0\) represents the strength (weight) of the directed edge from node \(j\) to node \(i\), and \(\mathbf{A}\) is referred to as the weighted adjacency matrix.
In undirected networks, the adjacency matrix is symmetric (\(a_{ij} = a_{ji}\)), so the weights are identical for both directions and the Laplacian matrix remains symmetric and positive semi-definite. 
In the case of weighted undirected networks, all matrix entries simply reflect the edge weights, with zero indicating absence of an edge. Conversely, for (un)weighted directed networks, the adjacency matrix is generally asymmetric, and the corresponding (weighted) Laplacian also becomes asymmetric, which significantly affects the spectral properties and hence the convergence behavior. 
Specifically, the presence of asymmetry or heterogeneous weights may lead to complex eigenvalues, and the standard tools for analyzing synchronization and consensus may require additional care to account for the weighted structure.

\begin{figure}
    \centering
    \includegraphics[width=1\linewidth]{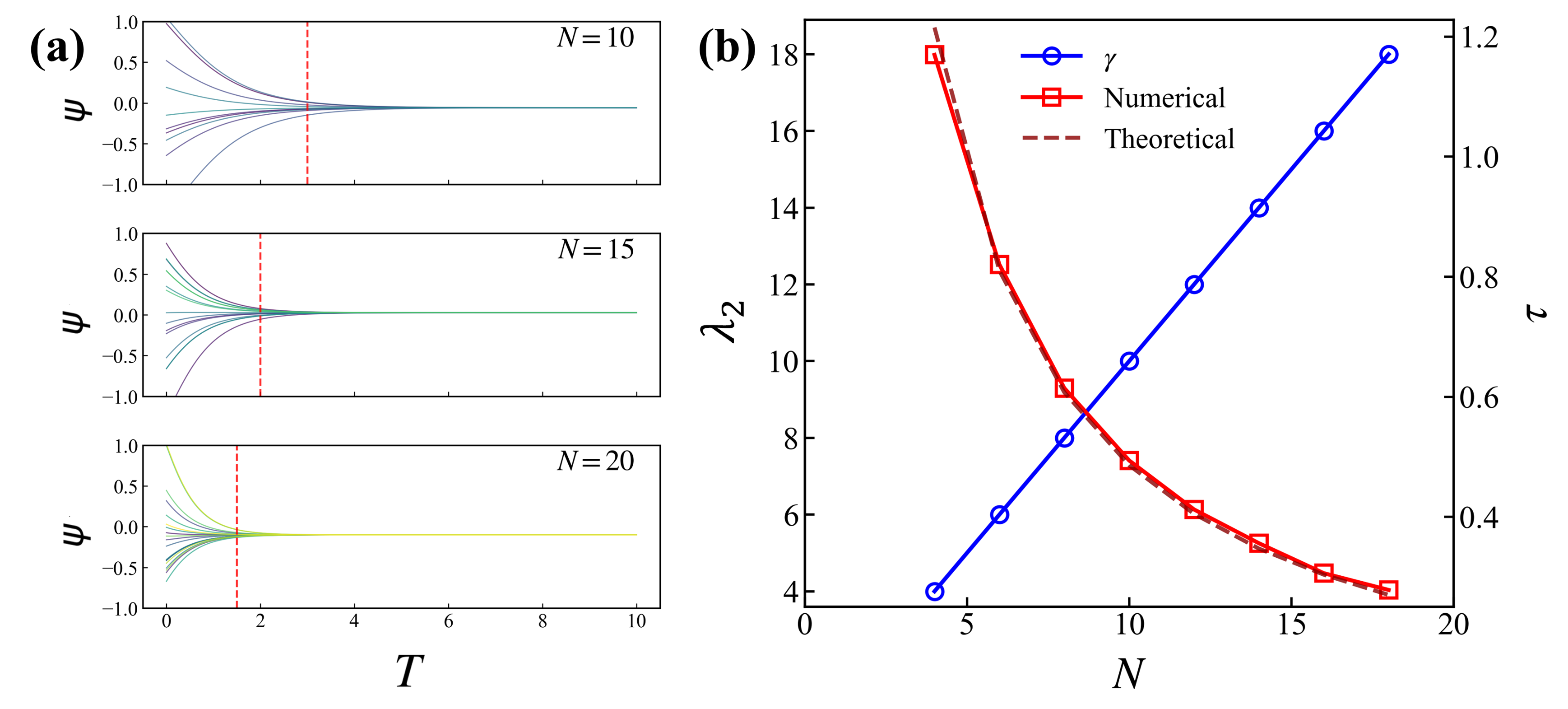}
    \caption{Synchronization dynamics in complete undirected networks. (a) Time evolution of the oscillator phases \(\psi\) in complete undirected graphs with \(N = 10, 15, 20\) nodes. All oscillators are governed by the Kuramoto dynamics with initial phases independently drawn from the uniform distribution \(\mathcal{U}(-1,1)\). The red dashed line indicates the time when the maximum phase difference drops below 0.01, marking the numerical convergence time.
        (b) Scaling of the minimal coupling strength \(\gamma\) (left axis, blue circles) and the convergence time constant \(\tau\) (right axis, red squares) as functions of network size \(N\). Numerical results are compared with theoretical predictions (brown dashed line) based on Eq.~\ref{con:undirected_convergence}.}
    \label{fig:generate}
\end{figure}

\subsection{Metric for Undirected Networks}

In a connected undirected network, the Laplacian matrix \(\lap\) is real, symmetric, and positive semi-definite. Its eigenvalues can be ordered as
\[
0 = \lambda_1 < \lambda_2 \le \cdots \le \lambda_n,
\]
with corresponding orthonormal eigenvectors \(\{\bm{u}_i\}_{i=1}^n\). The zero eigenvalue \(\lambda_1 = 0\) corresponds to the uniform (synchronized) mode, while the second-smallest eigenvalue \(\lambda_2\), known as the algebraic connectivity, governs the slowest convergence mode in the system.

Let the global state vector be \(\bm{\psi}(t) = [\psi_1(t)^\top, \dots, \psi_n(t)^\top]^\top\), where each \(\psi_i(t) \in \mathbb{R}^d\). Projecting the initial condition onto the Laplacian eigenbasis yields:
\[
\bm{\psi}(0) = \sum_{i=1}^n c_i\,\bm{u}_i,
\qquad
c_i = \bm{u}_i^\top \bm{\psi}(0).
\]
The linear coupling term \((\lap \otimes \mathbf{H})\,\bm{\psi}\) then gives rise to the modal evolution:
\begin{equation}
    \bm{\psi}(t) = c_1\,\bm{u}_1 + \sum_{i=2}^n c_i\,\exp\left(-\lambda_i t\right)\,\bm{u}_i,
    \label{eq:decomposition}
\end{equation}
where \(\mathbf{H}\) has been normalized into the time constant for simplicity. The first term represents the synchronized manifold, while the remaining terms convergence exponentially at rates determined by the eigenvalues \(\lambda_i\).

Define the synchronization error as
\[
\bm{\delta}(t) = \bm{\psi}(t) - c_1\,\bm{u}_1,
\]
and consider the Lyapunov function
\[
V(t) = \tfrac{1}{2}\|\bm{\delta}(t)\|_2^2.
\]
Differentiating along the dynamics 
\(\dot{\bm{\psi}} = F(\bm{\psi}) - (\lap \otimes \mathbf{H})\,\bm{\psi}\),
and neglecting higher-order nonlinearities in \(\bm{\delta}\), we obtain:
\begin{equation}
\begin{aligned}
    \dot{V}(t)
    &= \bm{\delta}^\top\,\dot{\bm{\delta}} 
    \approx -\,\bm{\delta}^\top\,(\lap \otimes \mathbf{H})\,\bm{\delta}
    \le -\,\lambda_{\min}(\mathbf{H})\,\lambda_2\,\|\bm{\delta}\|_2^2 
   \\& = -\,2\,\lambda_{\min}(\mathbf{H})\,\lambda_2\,V(t),
   \end{aligned}
\end{equation}
which holds for strongly diffusive coupling, i.e.,$ \lambda_{min}(H) > 0$. When H is singular, partial synchronization can still occur within the coupled subspace, yielding finite convergence times despite$ \lambda_{min}(H)=0$.  Integrating the inequality gives
\[
\|\bm{\delta}(t)\|_2 \le \|\bm{\delta}(0)\|_2\, \exp\left(-\lambda_{\min}(\mathbf{H})\,\lambda_2\,t\right),
\]
and thus the characteristic convergence time constant satisfies
\[
\tau \sim \frac{1}{\lambda_{\min}(\mathbf{H})\,\lambda_2}.
\]
This result shows that the synchronization rate depends jointly on the inner coupling strength \(\lambda_{\min}(\mathbf{H})\) and the spectral connectivity \(\lambda_2\) of the network. In practice, for a given physical system, the inner coupling matrix \(\mathbf{H}\) is typically determined by agent-level dynamics, making \(\lambda_{\min}(\mathbf{H})\) effectively constant. Hence, the convergence time \(\tau\) is primarily governed by the algebraic connectivity \(\lambda_2\), making it the key factor for global synchronization efficiency, that is 
\begin{equation}
    \tau \sim \frac{1}{\lambda_2}.
    \label{con:undirected_convergence}
\end{equation}

We analyze the synchronization dynamics of Kuramoto oscillators \cite{Kuramoto1975} on complete undirected networks, which we use here as a canonical phase-reduction framework linking nonlinear oscillator dynamics to Laplacian spectral properties that are central to our synchronization analysis.
\begin{equation}
    \frac{d\psi_i}{dt} = \omega_i + \kappa \sum_{j=1}^{N} A_{ij} \sin(\psi_j - \psi_i),
\label{eq:kuramoto}
\end{equation}
where $\psi_i$ denotes the phase of oscillator $i$, $\omega_i$ is its intrinsic frequency (set to zero in our analysis), $A_{ij}$ is the adjacency matrix, and $\kappa$ represents the coupling strength (fixed at unity in simulations). In a complete graph, every node interacts with all others, leading to maximal structural symmetry and homogeneous coupling among oscillators.

In the synchronized regime, the phase differences between oscillators are small, i.e., 
\(|\psi_j - \psi_i| \ll 1\) for all connected pairs $(i,j)$. Under this condition, the coupling term in Eq.~\eqref{eq:kuramoto} can be linearized using $\sin(\psi_j - \psi_i) \approx \psi_j - \psi_i$. The dynamics then reduce to a linear diffusion process governed by the graph Laplacian:
\begin{equation}
    \frac{d\bm{\psi}}{dt} = -\kappa \lap \bm{\psi},
\end{equation}
where $\lap$ is the combinatorial Laplacian of the complete graph. This linear approximation provides a direct link between the synchronization rate and the algebraic connectivity $\lambda_2$, which dictates the exponential convergence toward phase alignment.

As illustrated in Fig.~\ref{fig:generate}, numerical simulations confirm that the convergence time constant $\tau$ closely agrees with the theoretical prediction derived from Eq.~\eqref{con:undirected_convergence}, yielding $\tau = 1.62/\gamma$ in this configuration. Notably, the algebraic connectivity $\lambda_2$ grows linearly with the network size $N$, indicating that larger complete graphs synchronize more efficiently due to enhanced spectral connectivity.

\subsection{Metric for Directed Networks}

In directed networks, the Laplacian matrix \( \lap \) is generally asymmetric and may admit complex-valued eigenvalues. Despite this, the diffusive coupling structure ensures that all eigenvalues lie in the closed right-half complex plane, satisfying \( \operatorname{Re}(\lambda_i) \geq 0 \) for all \( i \). Ordering the eigenvalues by their real parts yields
\[
0 = \operatorname{Re}(\lambda_1) \leq \operatorname{Re}(\lambda_2) \leq \cdots \leq \operatorname{Re}(\lambda_n).
\]
The real part of the second smallest eigenvalue, denoted as \(\operatorname{Re}(\lambda_2) \), provides a lower bound on the convergence rate of perturbations orthogonal to the synchronization manifold. This q
antity has been discussed in prior works as a heuristic indicator of convergence rate in directed networks~\cite{olfati2004consensus, jadbabaie2003coordination, li2010consensus}. These studies typically highlight the empirical correlation between \(\operatorname{Re}(\lambda_2)\) and synchronization speed through simulations or qualitative arguments, but do not provide a rigorous mathematical characterization.

In contrast, the present work establishes a formal and exact relationship between the convergence time constant \( \tau \) and the spectral quantity \( \operatorname{Re}(\lambda_2) \) using Lyapunov analysis. Specifically, we prove that
\[
\tau \sim \frac{1}{\operatorname{Re}(\lambda_2)},
\]
Analogously, we consider the quadratic Lyapunov function \(
V(t) = \frac{1}{2} \|\bm{\delta}(t)\|_2^2\) and assuming that the dynamics remain close to the synchronization manifold, a linearized approximation of Eq.~\eqref{eq:dynamics} applies, leading to the differential inequality
\begin{equation}
    \dot{V}(t) \leq -\operatorname{Re}(\lambda_2)_{\mathrm{eff}} V(t),
    \label{eq:lyap_ineq}
\end{equation}
where \( \operatorname{Re}(\lambda_2)_{\mathrm{eff}} > 0 \) is an effective convergence rate. Integrating this inequality yields an exponential convergence bound:
\begin{equation}
    \|\bm{\delta}(t)\|_2 \leq \|\bm{\delta}(0)\|_2\, e^{-\operatorname{Re}(\lambda_2)_{\mathrm{eff}} t},
    \label{eq:decay}
\end{equation}
with convergence time constant \( \tau := 1/\operatorname{Re}(\lambda_2)_{\mathrm{eff}} \). While this expression captures the observable decay rate, it does not yet reveal its precise dependence on network structure.

To address this, consider the Lyapunov equation
\begin{equation}
    \lap^* \matP + \matP \lap = -\matI,
    \label{eq:lyap_eq}
\end{equation}
where \( \matP \succ 0 \) is the unique positive definite solution and $\lap^*$ denotes the transpose of the Laplacian matrix. The matrix \( \matP \) defines a weighted energy metric that governs the dissipation of perturbations in the linearized system. Under this formulation, the convergence time constant is given by \( \tau = \lambda_{\max}(\matP),\) providing an exact characterization of synchronization time based on the network's Laplacian spectrum and geometric energy dissipation.

This behavior is further captured by the variational form of the convergence time:
\begin{equation}
    \tau = \max_{\|\bm{v}\| = 1} \int_0^{\infty} \|e^{-\lap t} \bm{v}\|_2^2 \, dt,
    \label{eq:energy_integral}
\end{equation}
which corresponds to the worst-case integrated trajectory energy over time. An asymptotic expansion of this expression yields
\begin{equation}
    \tau = \frac{1}{\operatorname{Re}(\lambda_2)} + \frac{\|\lap - \lap^*\|_F}{2 \operatorname{Re}(\lambda_2)^2} + \mathcal{O}\left(\frac{1}{\operatorname{Re}(\lambda_2)^3}\right),
    \label{eq:asymptotic}
\end{equation}
where \( \|\cdot\|_F \) denotes the Frobenius norm. The second term quantifies the penalty due to non-normality, highlighting its role in delaying convergence beyond the baseline set by \( 1/\operatorname{Re}(\lambda_2) \).

Equally, we study the synchronization dynamics of Kuramoto model by Eq.~\eqref{eq:kuramoto} on directed ring networks of varying size \(N\). Each oscillator follows the standard Kuramoto model, with initial phases independently drawn from the uniform distribution \(\mathcal{U}(0,1)\). 

\begin{figure}
    \centering
    \includegraphics[width=1\linewidth]{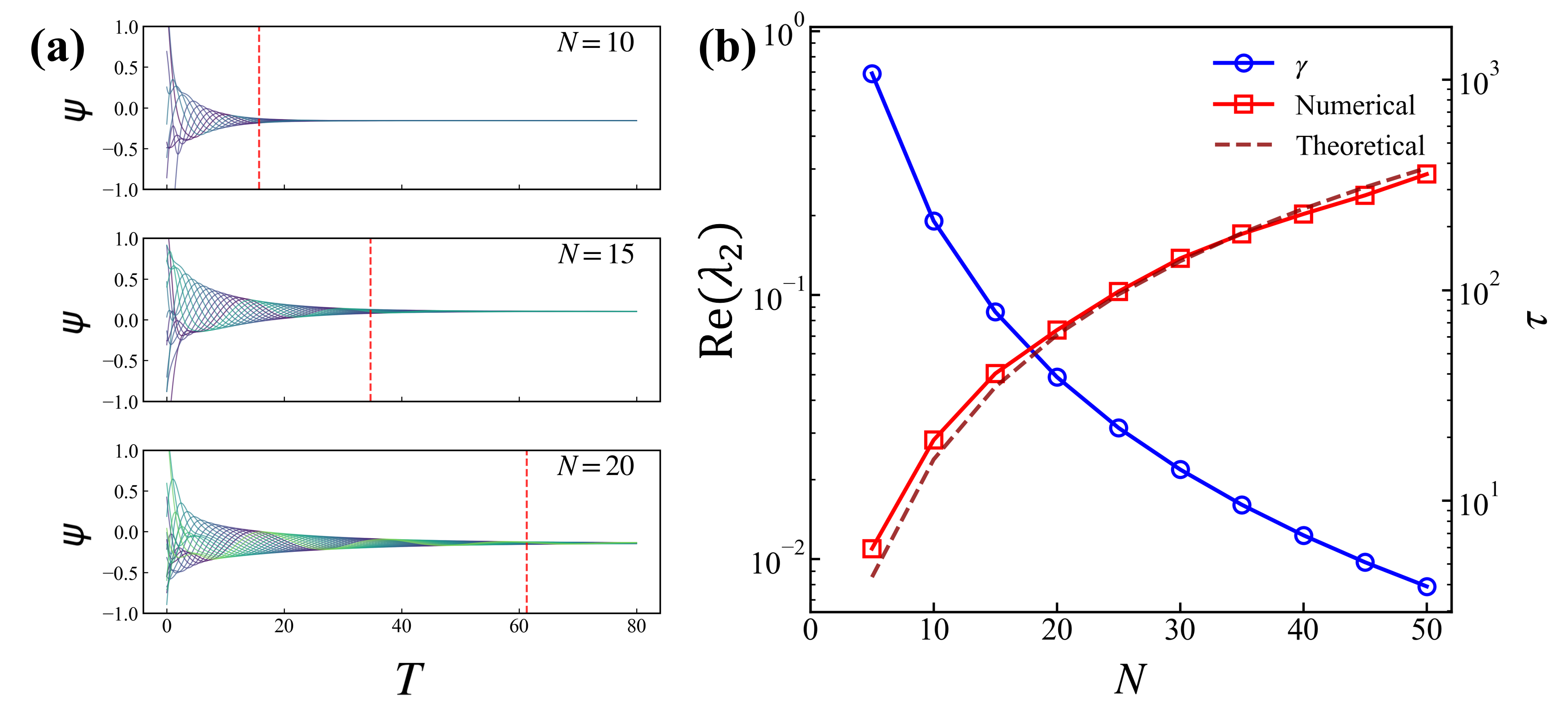}
  \caption{
        Synchronization dynamics and convergence time in directed ring networks. 
        (a) Time evolution of the oscillator phases \(\psi\) in directed ring networks with \(N = 10, 15, 20\) nodes. All oscillators start with initial phases drawn independently from a normal distribution \(\mathcal{N}(0,1)\). The system evolves under the Kuramoto dynamics. The red dashed line marks the moment when synchronization is numerically considered complete, defined as the time when the maximum phase difference among oscillators drops below 0.01. 
        (b) Dependence of the minimal coupling strength \(\operatorname{Re}(\lambda_2)\) (left axis, blue circles) required to achieve synchronization and the convergence time constant \(\tau\) (right axis, red squares) on the network size \(N\). Numerical results for \(\tau\) (red squares) are compared with theoretical predictions (dashed brown line) based on Eq.~\ref{eq:asymptotic}.
    }
    \label{fig:ring_sync}
\end{figure}

The numerical results presented in Fig.~\ref{fig:ring_sync} demonstrate excellent agreement with the theoretical prediction given by Eq.~\eqref{eq:asymptotic}, where \(\tau =3/\operatorname{Re}(\lambda_2)\) in this setting.

From a physical perspective, the difference in spectral properties between undirected and directed networks manifests clearly in the observed time series of phase evolution. In undirected networks shown in Fig.~\ref{fig:generate}(a), where all Laplacian eigenvalues are real and nonnegative, perturbations converge monotonically toward the synchronization manifold. This leads to smooth and gradual alignment of oscillator phases over time, with no oscillatory overshoot.

In contrast, directed networks typically exhibit complex Laplacian spectra, where eigenvalues can have significant imaginary components. As shown in Fig.~\ref{fig:ring_sync}(a), this leads to transient oscillatory behaviors in the phase trajectories before synchronization is achieved. The real part of each eigenvalue governs the exponential convergence rate of the corresponding mode, while the imaginary part introduces rotational or oscillatory dynamics. Thus, \(\operatorname{Re}(\lambda_2)\) determines how fast the system approaches synchrony, while \(\operatorname{Im}(\lambda_2)\) modulates the presence and frequency of transient oscillations, while primarily governs transient oscillations and does not alter asymptotic convergence.

These distinctions highlight the nuanced roles of complex eigenvalues in synchronization dynamics: real parts regulate convergence, while imaginary parts induce transient coherence waves. Understanding both components is essential for interpreting the full dynamical landscape of directed systems.

Across both undirected and directed networks, the convergence time constant~\(\tau\) serves as a fundamental indicator of global synchronization efficiency. 
In undirected graphs, where the Laplacian matrix is symmetric and its spectrum real, \(\tau\) is directly governed by the algebraic connectivity~\(\lambda_2\) as established in Eq.~\eqref{con:undirected_convergence}. 

For directed networks, the Laplacian spectrum generally lies in the complex plane. 
Nevertheless, theoretical analysis shows that the real part of the smallest nonzero eigenvalue, \(\operatorname{Re}(\lambda_2)\), remains a reliable predictor of the asymptotic convergence rate, as demonstrated by the Lyapunov-based bounds and energy dissipation relation expressed in Eq.~\eqref{eq:asymptotic}. 
This quantity effectively extends the notion of algebraic connectivity to non-Hermitian network topologies, capturing the dominant decay mode that governs synchronization dynamics.

The dependence of the convergence rate on \(\operatorname{Re}(\lambda_2)\) is consistent with the established theoretical understanding of directed complex networks, where the effective algebraic connectivity under balanced or weakly non-normal conditions can be approximated by the real component of the smallest nonzero Laplacian eigenvalue~\cite{Wu2005,Wu2007,ImaeCai2018}. 

Accordingly, the generalized Fiedler value is defined as
\[
\gamma = \operatorname{Re}(\lambda_2) \sim \frac{1}{\tau},
\]
representing an effective spectral gap that determines the exponential synchronization rate, which provides a coherent approach for comparing synchronization performance across both symmetric and asymmetric network topologies.

\section{FGIA  Approach}
Now we come to consider edge removal specific form by perturbing the Laplacian matrix. Given the intrinsic relationship between network topology and the performance of synchronization \cite{liu2022network,Gao2016Universal}, the removal of edges invariably increases the system convergence time, thereby reducing synchronization performance \cite{jiang2023searching}. To clearly illustrate the impact of edge attacks on both network connectivity and system dynamics, we apply identical state perturbations across the systems and measure the corresponding convergence times, as shown in Fig.~\ref{fig:generate}(a-c). Collectively, these results confirm that edge removal compromises network connectivity, thereby altering the dynamic characteristics of the system.

From a spectral perspective, the impact of edge removal can be understood through its effect on the Laplacian eigenmodes that support collective synchronization dynamics. In highly symmetric networks, structural regularity is encoded in the Laplacian operator through degenerate or near-degenerate eigenmodes, which define invariant subspaces associated with synchronization and other coherent collective behaviors. Consequently, synchronization stability is not governed solely by spectral gaps, but is fundamentally intertwined with the underlying network symmetries reflected in the Laplacian spectrum.

Structural perturbations induced by edge removal alter these spectral properties by lifting degeneracies and reshaping the low-order eigenmodes that support symmetry-protected invariant subspaces. As a result, collective patterns sustained by network symmetry—such as synchronized states or other coherent phase-organized modes—can be destabilized even when global connectivity is preserved. Within the proposed FGIA framework, such symmetry breaking is not imposed explicitly. Instead, it emerges implicitly through targeted perturbations of the Laplacian spectrum, whereby edges that contribute most strongly to the spectral support of synchronization-related modes are selectively removed, leading to a systematic suppression of coherent dynamics.

Since for the Laplacian matrix an undirected graph can be regarded as a special case of a directed graph,
we begin by considering the more general setting of directed networks.
Treating edge removal as a perturbation to the Laplacian operator,
the perturbed eigenvalue problem can be written as
\begin{equation}
    (\lap + \Delta\lap)(\vect{x} + \delta\vect{x})
    = (\gamma + \delta\gamma)(\vect{x} + \delta\vect{x}),
\end{equation}
where $\Delta \lap$ denotes the perturbation matrix induced by edge removal,
while $\delta\gamma$ and $\delta\vect{x}$ represent the corresponding changes
in the eigenvalue and eigenvector of $\lap$, respectively.
For the removal of a directed edge $e_{i,j}$,
the Laplacian perturbation takes the rank-one form
\begin{equation}
    \Delta\lap(e_{i,j})
    = -(\vect{e}_i - \vect{e}_j)(\vect{e}_i - \vect{e}_j)^\top,
\end{equation}
where $\vect{e}_i$ is the standard basis vector.
The perturbed eigenvector $\delta\vect{x}$ can be expanded in the basis of right eigenvectors of $\lap$ as
\begin{equation}
    \delta\vect{x}
    = \sum_{k\neq2} c_k \vect{x}_k,
    \label{eq:eigen_expansion}
\end{equation}
where $\vect{x}_k$ is the $k$-th right eigenvector satisfying
$\lap\vect{x}_k = \lambda_k\vect{x}_k$.

Expanding the eigenvalue equation to first order in the perturbation and neglecting
terms of $\mathcal{O}(\|\Delta\lap\|^2)$,
one obtains
\begin{equation}
    \lap\,\delta\vect{x}
    + \Delta\lap\,\vect{x}
    = \gamma\,\delta\vect{x}
    + \delta\gamma\,\vect{x}.
\end{equation}
Left-multiplying by $\vect{y}^\top$, where $\vect{y}$ denotes the left eigenvector
associated with $\gamma$ and satisfying $\vect{y}^\top\lap = \gamma\vect{y}^\top$,
and using the normalization condition $\vect{y}^\top\vect{x} = 1$,
the eigenvalue shift can be expressed as
\begin{equation}
    \delta\gamma
    = \vect{y}^\top \Delta\lap \vect{x}
    + \vect{y}^\top \Delta\lap \delta\vect{x}.
\end{equation}

Substituting the eigenvector expansion in Eq.~\eqref{eq:eigen_expansion} into the second term
and using the bi-orthogonality condition
$\vect{y}_k^\top \vect{x}_\ell = \delta_{k\ell}$,
the expansion coefficients are given by
\begin{equation}
    c_k
    = \frac{\vect{y}_k^\top \Delta\lap \vect{x}}{\gamma - \lambda_k},
    \qquad k \neq 2.
\end{equation}

Collecting all contributions, the perturbation of the eigenvalue $\gamma$
induced by the removal of edge $e_{i,j}$ in a directed network can be written as
\begin{align}
    \delta\gamma(e_{i,j})
    &= -(\nabla \epsilon_{ij})^2 \nonumber \\
    &\quad - \mathrm{Re}\!\left[
        \sum_{k>2}
        \frac{
            \vect{y}^\top \Delta\lap \vect{x}_k \;
            \vect{y}_k^\top \Delta\lap \vect{x}
        }{
            \gamma - \lambda_k
        }
    \right],
    \label{eq:full_perturbation}
\end{align}
where the negative sign reflects the fact that edge removal weakens network connectivity.
As a result, the real part of the second smallest eigenvalue $\mathrm{Re}(\lambda_2)$ decreases,
implying $\delta\gamma < 0$ and hence a larger synchronization time scale
$\tau \sim 1/\gamma$.
This behavior is consistent with the physical intuition that structural degradation
suppresses collective synchronization.

In this work, we define the Fiedler gradient to quantify the local spectral contribution of an edge to the overall synchronization dynamics, based on the sensitivity of the algebraic connectivity to edge modifications. This gradient is motivated by classical perturbation theory, which explores how small changes in network structure can affect the Fiedler value\cite{fiedler1973algebraic}.
Thus, the leading-order contribution $(\nabla \epsilon_{ij})^2$ is determined by the squared
Fiedler gradient:
\begin{equation}
    \nabla \epsilon_{ij}
    = \mathrm{Re}\!\left( \big| (y_i - y_j)\, x_j \big| \right),
\label{eq:gradient_general}
\end{equation}
where $x_j$ denotes the $j$-th component of the normalized right eigenvector $\vect{x}$
associated with $\gamma$, and $y_i$, $y_j$ are the corresponding components of the left
eigenvector $\vect{y}$.

The vertical bars $|\cdot|$ indicate the complex modulus,
while $\mathrm{Re}(\cdot)$ extracts the real part.
The summation over higher-order modes in Eq.~\eqref{eq:full_perturbation}
represents the second-order spectral correction.
Physically, the gradient $\nabla \epsilon_{ij}$ quantifies the local misalignment
between the left and right Fiedler eigenvectors across edge $(i,j)$,
providing a direct measure of that edge’s contribution to information propagation
efficiency and dynamical coupling strength within the network.

In the case of undirected networks, the Laplacian matrix $\lap$ is symmetric and
positive semi-definite, with real eigenvalues and orthonormal eigenvectors.
In this symmetric setting, the left and right eigenvectors coincide
($\vect{y}=\vect{x}$), rendering the biorthogonal correction unnecessary and causing
the complex modulus and real-part operators in Eq.~\eqref{eq:gradient_general} to collapse to a purely real
expression. Thus, the eigenvalue perturbation due to edge $(i,j)$ therefore simplifies to
\begin{equation}
    \nabla \epsilon_{ij} = (x_i - x_j)^2,
    \label{eq:full_perturbation_undirected}
\end{equation}
where $\vect{x}$ is the Fiedler eigenvector corresponding to the algebraic connectivity
$\lambda_2$.
The leading term $(x_i - x_j)^2$ precisely quantifies the change in $\lambda_2$ due to
the structural modification on edge $(i,j)$, offering a clear interpretation:
edges spanning high eigenvector gradient regions contribute more significantly to
global synchronization and network robustness.

Note that all derivations and computations presented above do not require the Laplacian matrix entries to be binary (i.e., strictly 0 or 1). Instead, the approach naturally extends to weighted networks where edge weights can take arbitrary non-negative values. This generality implies that the perturbation analysis, the definition of the Fiedler gradient, and the subsequent edge ranking procedures remain valid regardless of whether the network is unweighted or weighted. Consequently, the proposed method is applicable to any static network topology, encompassing a broad class of real-world systems characterized by heterogeneous interaction strengths. Such flexibility significantly enhances the practical relevance of the approach in analyzing and controlling synchronization phenomena in complex networks with diverse weighted structures.

Moreover, since eigenvectors in the undirected case are orthogonal and real, the second-order term remains suppressed under small perturbations. This yields a consistent conclusion across both directed and undirected settings: The edge gradient is the primary determinant of spectral sensitivity.

This universality validates the use of Fiedler gradient-based heuristics in ranking edge importance for control and design in complex networks.

Accordingly, we can obtain the objective function for optimal edge removal as 
\begin{equation}
    Tar(i,j) = \arg\max_{i,j} \nabla \epsilon_{ij},
    \label{eq:target}
\end{equation}
which identifies optimal edge attacks under connectivity constraints via spectral perturbation theory. By maximizing the nodal gradient $\nabla \epsilon_{ij}$ in the Fiedler vector $\mathbf{x}$, it targets edges bridging the two dynamically heterogeneous modules (positive/negative components of $\mathbf{x}$). Such edges critically mediate inter-module coordination, where removal maximally degrades $\gamma$ to suppress synchronization, while preserving global connectivity as these edges are typically non-bridge connections.

\begin{figure}
    \centering
    \includegraphics[width=1\linewidth]{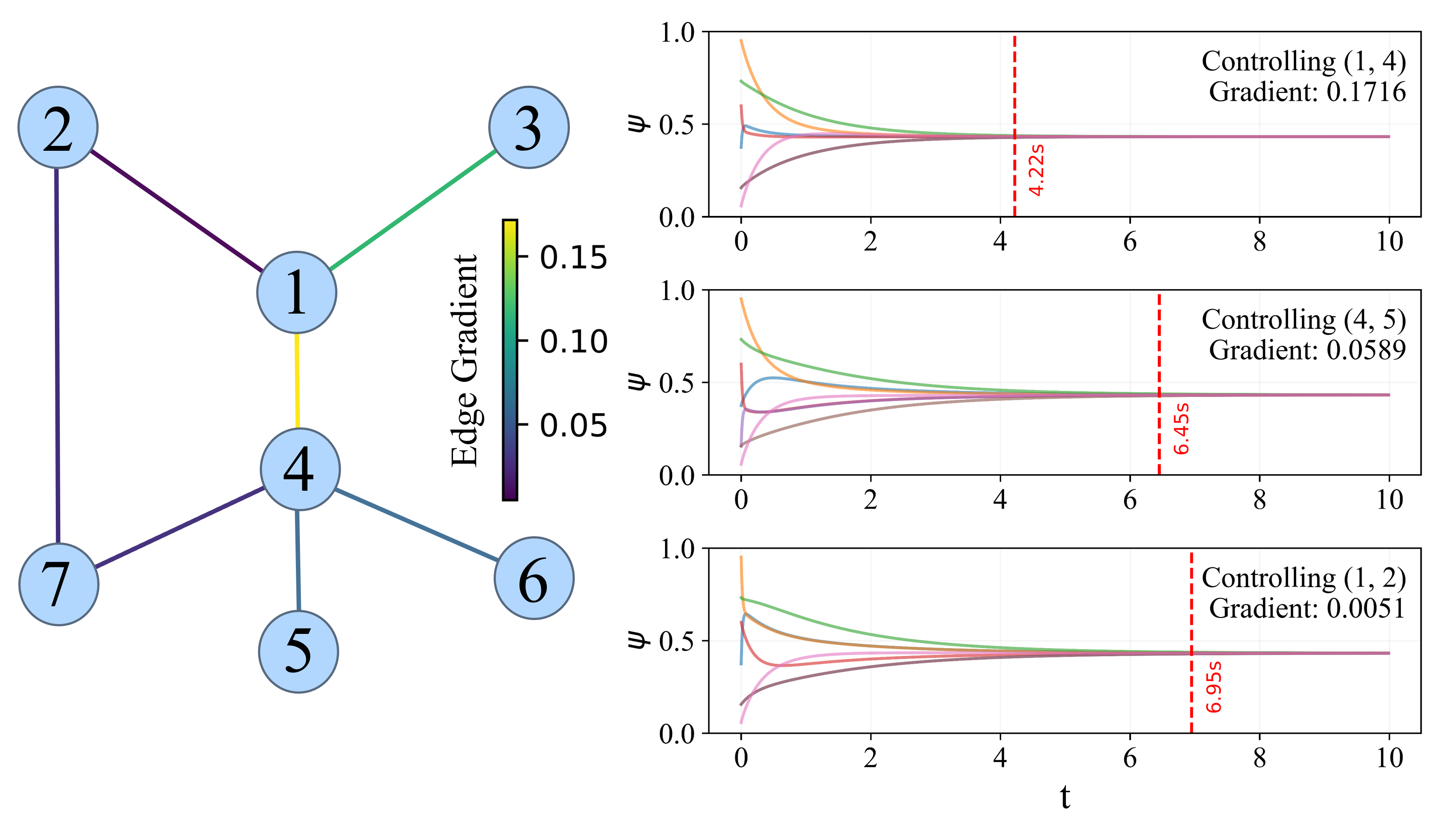}
    \caption{Temporal evolution of node phases $\psi_i$ ($1 \leq i \leq 7$) under control of three representative edges selected based on their Fiedler gradient in the example network. Edge colors denote gradient magnitudes. In the right panel, red dotted lines indicate the time at which the system reaches synchronization, marking the threshold used to determine the convergence time.}

    \label{fig:controlling}
\end{figure}

To validate this principle,  we compare convergence timelines under three distinct interventions in Fig.~\ref{fig:controlling}: edge-specific coupling enhancements for the maximum-gradient edge $(1,4)$, median-gradient edge $(4,5)$, and minimum-gradient edge $(1,2)$. The observed convergence time exhibit an inverse correlation with edge gradient magnitudes -- the sequence of convergence times aligns precisely with the descending order of gradient values ($\delta \gamma^{(1,4)} > \delta \gamma^{(4,5)} > \delta \gamma^{(1,2)}$). The results demonstrate that gradient-driven edge prioritization effectively captures structural determinants of synchronization.

We further elucidate this result from the perspective of network community structure. The pronounced sensitivity of inter-community edges can be attributed to the steep Fiedler gradient values observed in the distribution of the Fiedler vector.

In this work, communities are identified through a \textit{hierarchical spectral bisection} procedure guided by the Fiedler vector. At each bisection step, nodes are partitioned into two subsets according to the sign of their Fiedler vector components: nodes with nonnegative components are assigned to $C^{+}$, while those with negative components are assigned to $C^{-}$. This recursive process proceeds up to the maximum depth $d_{\max}$, thereby capturing increasingly fine-grained modular structures within the network. Under this approach, intra-community edges are defined as edges connecting nodes that share the same sign at a given partitioning level, whereas inter-community edges are those that cross between nodes of opposite signs at any bisection step, thereby linking distinct spectral modules. Importantly, this classification is dynamic: after each edge removal in the FGIA, the Fiedler vector is recomputed, resulting in an updated community structure and a reclassification of intra- and inter-community edges.

\begin{figure}
    \centering
    \includegraphics[width=0.8\linewidth]{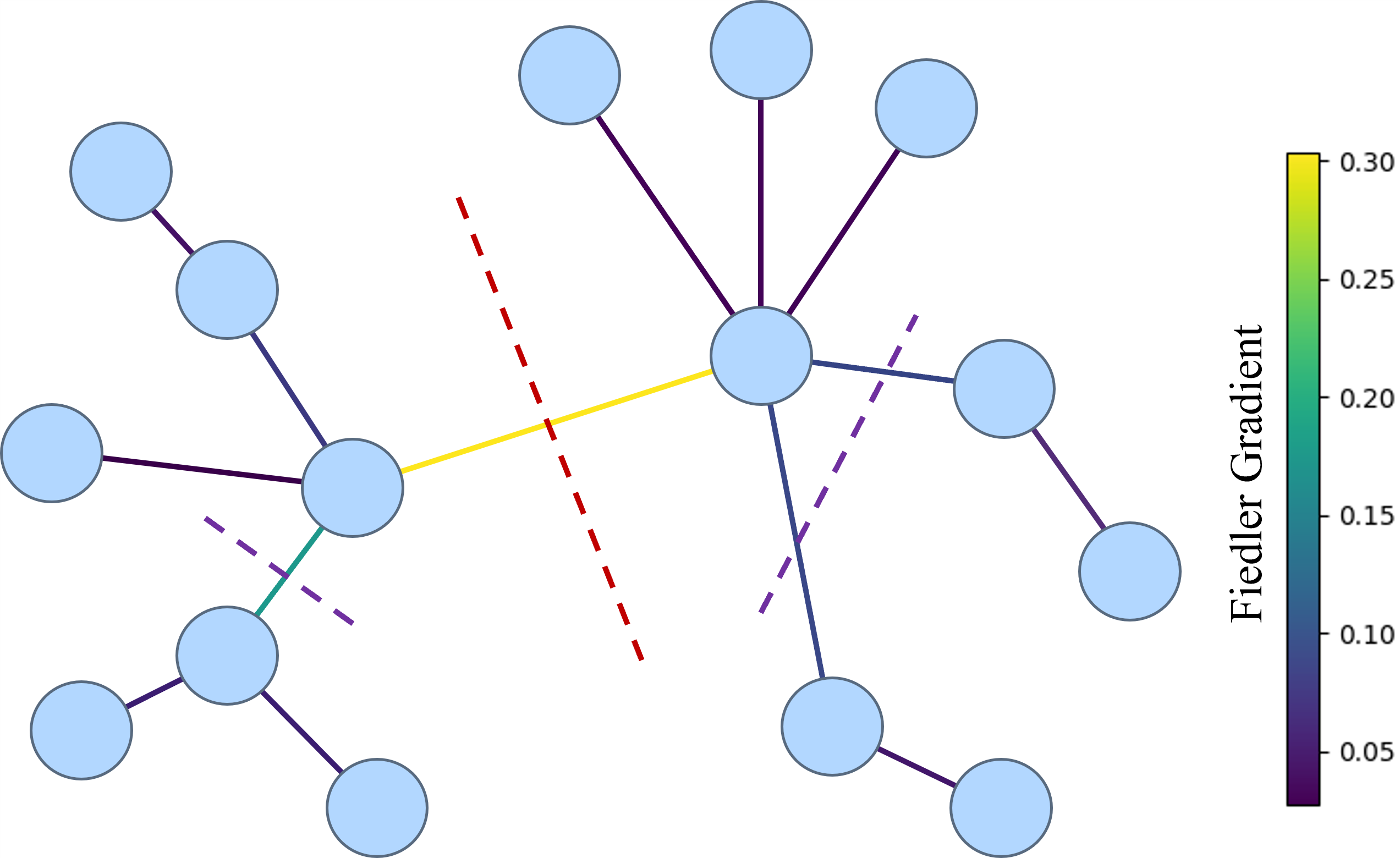}
    \caption{Diagram of community partition and the Fiedler gradient. a demonstrative case  in a Barabási–Albert (BA) network with $15$ nodes and  attachment parameter $m=1$. The color bar shows the Fiedler gradient. The red dotted line represents the first partition, and the purple dotted lines represent the second partition.}
    \label{fig:community}
\end{figure}

As illustrated in Fig.~\ref{fig:community}, these bridging edges connect nodes with markedly different eigenvector components. This phenomenon arises from three primary factors. First, structural heterogeneity causes inter-community edges to span regions with distinct local connectivity patterns, producing sharp transitions in the spectral embedding space. Second, the intrinsic spectral properties of the Fiedler vector naturally partition the network, thereby amplifying gradient magnitudes at community boundaries. Third, in the resistive network analogy, bridging edges carry a disproportionately large current, as quantified by Eq.~\eqref{eq:full_perturbation}.

Consequently, this disparity in Fiedler gradient values accounts for the asymmetric impact of edge removal: eliminating high-gradient inter-community edges disrupts the spectral bisection balance, substantially reducing $\gamma$ and prolonging convergence. In contrast, intra-community edges, which display relatively uniform gradient values, contribute minimally to $\gamma$, and their removal has a correspondingly limited effect on synchronization.

Therefore, we can degrade $\gamma$ of the network efficiently by calculating the edge with the maximum gradient as the optimal attack target . In many cases, however, we must ensure that the network is connected, while this using Eq.~\eqref{eq:target} cannot guarantee the connectivity of the network. Therefore, an optimized ranking edge removal algorithm of the Fiedler Gradient Iterative Attack (FGIA)  is proposed to ensure the edge optimal attack strategy of the network in the case of connectivity. 
\begin{algorithm}
\caption{Fiedler Gradient Iterative Attack (FGIA) with Fallback Strategy}
\label{alg:FGIA}
\SetKwInOut{Input}{Input}
\SetKwInOut{Output}{Output}

\Input{Network $G=(V,E)$, attack budget $k$, bisection depth $d_{\text{max}}$}
\Output{Attack edge sequence $E_{\text{attack}}$}
\BlankLine

\textbf{Initialize}: $E_{\text{attack}} \leftarrow \emptyset$, $G^{(0)} \leftarrow G$ \\
\For{$t \leftarrow 0$ \KwTo $k-1$}{
  \textbf{Step 1}: Compute Laplacian of $G^{(t)}$ \\
  \textbf{Step 2}: Solve Fiedler eigenpair $(\gamma^{(t)}, \vect{x}^{(t)})$ \\
  \textbf{Step 3}: Perform hierarchical spectral bisection up to depth $d_{\text{max}}$ \\
  \textbf{Step 4}: Compute Fiedler gradient $\nabla \epsilon_{ij}$ for cross edges \\
  \textbf{Step 5}: Detect bridge edges using Tarjan’s algorithm \\
  \textbf{Step 6}: Filter out bridge edges to form candidate set $E_{\text{non-bridge}}$ \\
  \textbf{Step 7}: Rank candidate edges by $\delta\gamma$ (descending order) \\
  \textbf{Step 8}: Validate connectivity iteratively; \\
  \Indp
    \If{$E_{\text{non-bridge}} = \emptyset$}{
      Apply fallback: include community-internal edges or next-best edges
    }
  \Indm
  \textbf{Step 9}: Remove selected edge $e^*$, update $E_{\text{attack}}$ and graph $G^{(t+1)}$
}
\Return $E_{\text{attack}}$
\end{algorithm}

As detailed in Alg.~\ref{alg:FGIA}, FGIA suppresses network synchronization by iteratively removing edges while preserving global connectivity. The procedure begins with constructing the Laplacian matrix $\lap$ and computing its Fiedler vector $\bm{x}$, corresponding to the eigenvector associated with the second smallest eigenvalue $\lambda_2$ of $\lap$. This vector encodes crucial information about the network’s connectivity and synchronization dynamics.

At each iteration, up to the predefined attack budget $k$, nodes are recursively partitioned into two communities based on the sign of their Fiedler vector components: $C_+ = \{ i \mid x_i \geq 0 \}$ and $C_- = \{ i \mid x_i < 0 \}$. This partitioning is performed hierarchically up to a maximum depth $d_{\text{max}}$, where deeper levels yield finer-grained community structures. The set of cross-community edges $E_{\mathrm{cross}}$ connecting nodes between $C_+$ and $C_-$ (or between subsequent subdivisions at deeper levels) is identified as the primary candidate set for removal.

To ensure network connectivity, Tarjan’s algorithm~\cite{Tarjan72Depth} is applied to detect bridge edges—whose removal would disconnect the network—and these are excluded from the candidate set. Each remaining candidate edge $(i,j)$ is then assigned a Fiedler gradient value, $ \nabla \epsilon_{ij} = (x_i - x_j)^2$ for undirected networks (or $\mathrm{Re}\big|(y_i - y_j)x_j\big|$ for directed networks), which quantifies its impact on synchronization. Edges with larger gradients are considered more critical; they are sorted in descending order to form an ordered list $S_{\mathrm{edge}}$, and the highest-ranked edge that preserves connectivity is permanently removed.

After each removal, the Laplacian $\lap$ is updated to reflect the modified topology, and the Fiedler vector $\bm{x}$ is recomputed using the inverse power method, preparing the network for the next iteration. This process repeats until the attack budget $k$ is exhausted, progressively dismantling the network’s synchronization structure while maintaining its overall connectivity.

\begin{figure}
    \centering
    \includegraphics[width=0.75\linewidth]{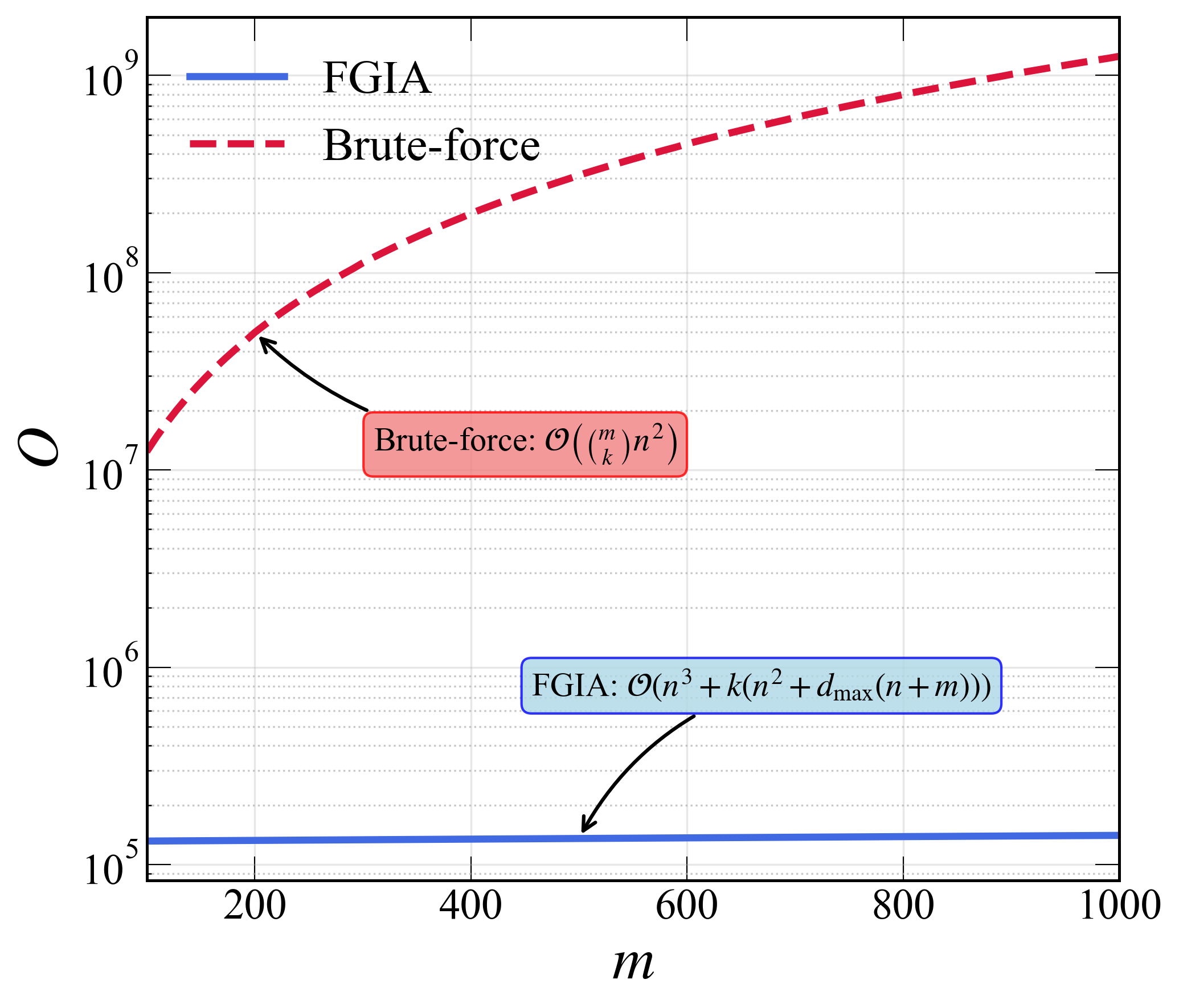}
    \caption{Comparison of the computational complexity between FGIA and brute-force algorithms. While the brute-force method grows combinatorially with the number of edges $m$, leading to an exponential increase in cost, the FGIA maintains polynomial-time scalability, making it practical for large-scale networks. With parameters $N=50,~k=2,~d_{max}=5$.}
    \label{fig:Complexity}
\end{figure}

\begin{figure}[!h]
    \centering
    \includegraphics[width=0.75\linewidth]{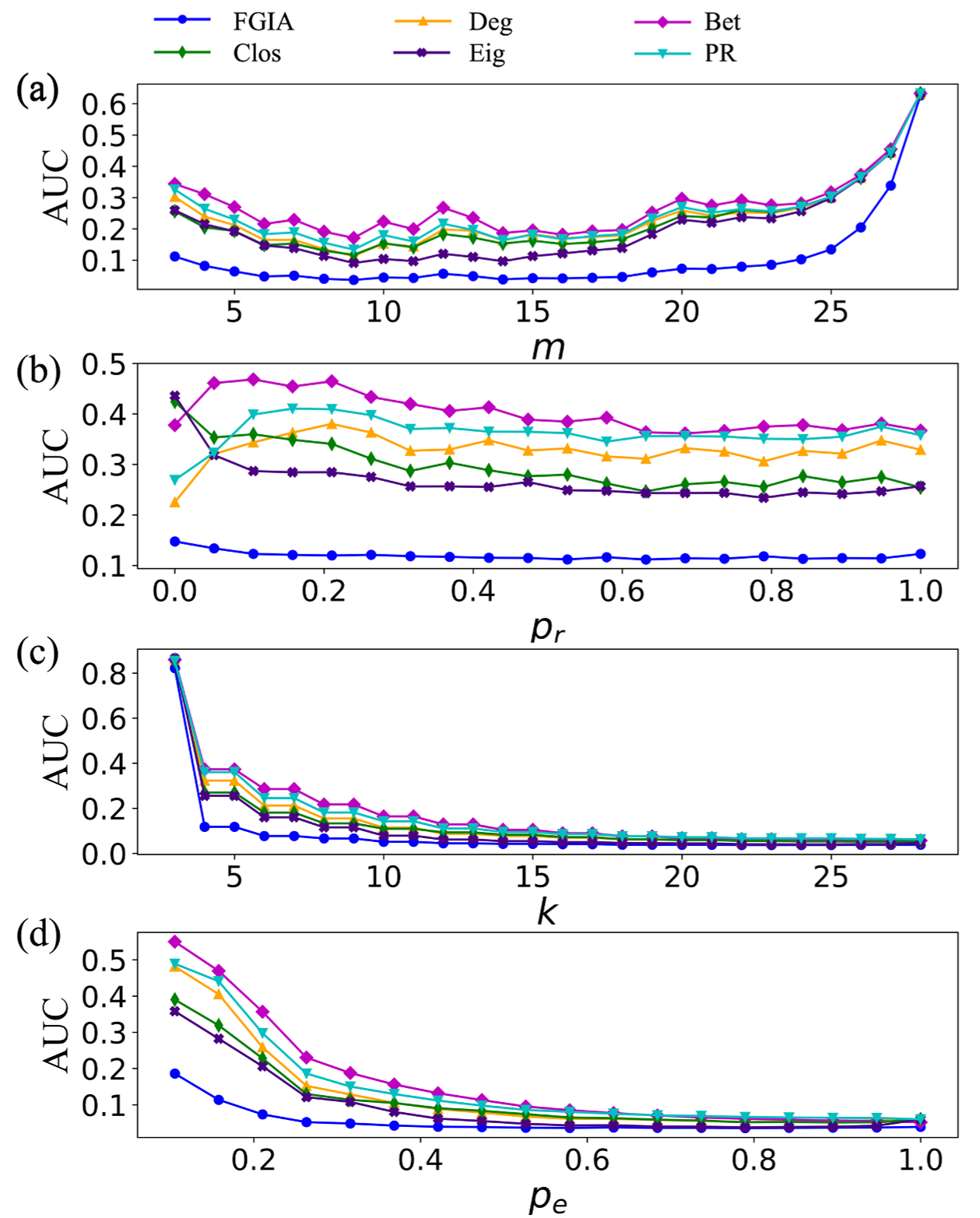}
    \caption{The AUC values obtained from several representative attack algorithms across different synthetic network models. (a) In Barabási–Albert (BA) scale-free networks, AUC values are plotted against the network growth parameter \( m \). (b) and (c) show the results for Watts–Strogatz (WS) small-world networks: in (b), AUC values vary with the rewiring probability \( p_r \) (with average degree fixed at \( k = 5 \)), and in (c), with \( k \) (with fixed \( p_r = 0.3 \)). (d) presents the results for Erdős–Rényi (ER) random networks, with AUC values shown as a function of the edge connection probability \( p_e \). Different colors represent different attack algorithms, as indicated in the legend.
}
    \label{fig:scanning}
\end{figure}

The computational efficiency of FGIA stems from the integration of spectral and graph-theoretic optimizations. Consider a network with $n$ nodes and $m$ edges, and let the attack budget $k$ denote the maximum number of edges to be removed. Unless otherwise specified, $k$ is chosen such that the iterative removals reduce the network to a minimally connected state: after removing $k$ edges, the remaining graph remains connected, but the deletion of any additional edge would disconnect it. This criterion ensures that the final attacked configuration is minimally connected and provides a natural termination condition for synchronization dismantling.

\begin{figure*}
    \centering
    \includegraphics[width=1\linewidth]{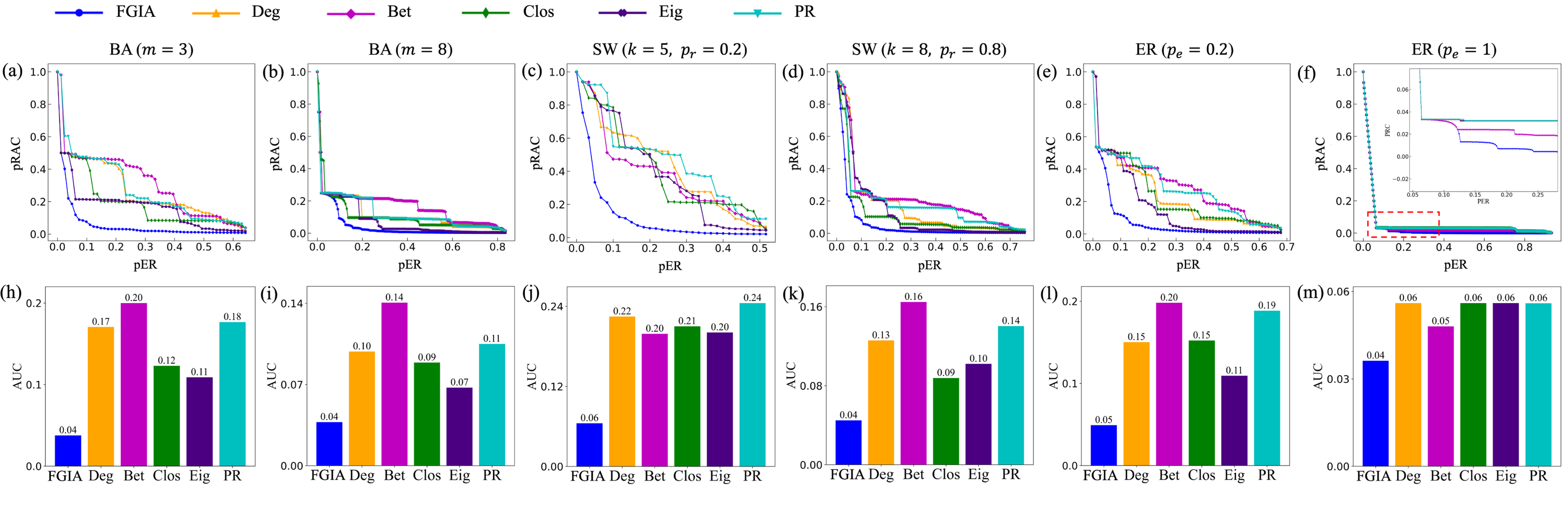}
    \caption{
    Edge-removal performance on undirected synthetic networks.
    Panels (a--f) show the percentage residual algebraic connectivity (pRAC) as a function of the percentage of edges removed (pER) for BA, WS, and ER networks with different parameters.
    Panels (g--l) report the corresponding AUC values.
    }
    \label{fig:attack_algorithms}
\end{figure*}

The algorithm begins with an initial eigen-decomposition of the Laplacian matrix, which incurs a computational cost of $\mathcal{O}(n^3)$. Subsequent iterations update the Fiedler vector via the inverse power method, contributing an additional $\mathcal{O}(k n^2)$ cost. At each iteration, a hierarchical spectral bisection is performed up to depth $d_{\textnormal{max}}$, introducing a further complexity of $\mathcal{O}(k\, d_{\textnormal{max}}\, m)$. Bridge detection, implemented via Tarjan’s algorithm~\cite{Tarjan72Depth}, operates in linear time with respect to the number of edges, resulting in $\mathcal{O}(k m)$ complexity over all iterations. Combining these components, the overall computational complexity of FGIA is
\[
\mathcal{O}\!\left(n^3 + k\big(n^2 + d_{\textnormal{max}}\, m\big)\right).
\]

For comparison, a brute-force approach would require evaluating all $\binom{m}{k}$ possible edge-removal combinations, with each combination necessitating an eigenvalue computation at cost $\mathcal{O}(n^2)$. This yields a prohibitive total complexity of
\[
\mathcal{O}\!\left(\binom{m}{k}\, n^2\right).
\]
In the limiting case of a complete graph, FGIA retains a polynomial complexity of $\mathcal{O}(n^3)$. By contrast, even for a relatively small $k=2$, the brute-force strategy escalates to $\mathcal{O}(n^6)$, underscoring the significant scalability advantage of FGIA.

We illustrate this contrast in computational growth in Fig.~\ref{fig:Complexity}. For instance, with $n=50$, $k=2$, and $d_{\textnormal{max}}=5$, the brute-force method becomes computationally infeasible when the number of edges reaches $m=1000$, requiring on the order of $10^9$ operations. FGIA, by leveraging spectral perturbation and efficient community-based edge filtering, maintains polynomial-time complexity and thus remains practical for larger and denser networks. In order to better compare with other algorithms and thereby demonstrate the advantages of FGIA, unless otherwise stated, we adopt $k=k_{max}$ at its maximum feasible value (driving the graph to minimal connectivity) and set $d_{\textnormal{max}}=1$, corresponding to single-edge removal per iteration.

Notice that, in Fig.~\ref{fig:attack_algorithms}(a-f), when the value of pER exceeds a certain threshold, the FGIA may exhibit a noticeable change in the AUC degradation rate. It indicates that continuing to remove edges based on the criterion defined in Eq.~\eqref{eq:target} may result in network disconnection. To prevent this, Alg.~\ref{alg:FGIA} adopts a degradation strategy based on the Fiedler gradient sequence of edges, thereby preserving the network's connectivity. Although this modification introduces additional computational complexity, we argue that ensuring connectivity while assessing network synchronization is of significant value.

However, if users opt for an attack strategy that does not require preserving network connectivity based on the network's functionality and attack objectives, then the bridge detection (Step 5), edge filtering (Step 6), and connectivity verification loop (Step 8) in Algorithm 1 can be omitted. Moreover, by modifying the ordering constraints (Step 7) so that the sorting extends from being applied solely to bridge edges to covering all inter-partition edges, we obtain a degenerate version of Alg.~\ref{alg:FGIA}. Although this version does not guarantee network connectivity, it is more destructive and achieves improved efficiency.

\section{Simulation and Analysis}

We evaluate the performance of the proposed FGIA method through extensive numerical experiments on synthetic networks, with the goal of assessing its effectiveness and robustness under diverse structural conditions. The analysis is organized along increasing levels of network complexity, starting from undirected networks, extending to weighted networks, and finally considering directed networks with asymmetric interactions. Throughout, FGIA is compared against conventional edge-removal strategies based on degree (Deg), betweenness centrality (Bet), closeness centrality (Clos), eigenvector centrality (Eig), and PageRank (PR).

We first consider undirected synthetic networks generated from Barabási–Albert (BA), Erdős–Rényi (ER), and Watts–Strogatz (WS) models with different parameter settings. These networks span heterogeneous scale-free structures, homogeneous random graphs, and small-world topologies with tunable clustering and randomness. Network functionality is quantified by the algebraic connectivity $\gamma$, normalized as the percentage residual algebraic connectivity (pRAC), which is tracked as a function of the percentage of edges removed (pER). In addition, the area under the pRAC--pER curve (AUC) is used as a scalar indicator of overall suppression efficiency, with smaller AUC values corresponding to more effective functional disruption.

Across all undirected network models, FGIA consistently induces the most rapid decay of pRAC with increasing pER, as summarized in Fig.~\ref{fig:attack_algorithms}. While centrality-based heuristics may exhibit competitive performance in specific regimes—such as closeness-based removal in certain small-world networks or degree-based strategies in moderately heterogeneous graphs—their effectiveness varies strongly with network topology. In contrast, FGIA maintains stable performance across all tested networks, achieving near-complete suppression of algebraic connectivity after removing only a small fraction of edges, typically on the order of $5\%$–$10\%$. This behavior highlights the advantage of FGIA’s gradient-driven spectral formulation, which directly targets edges that are functionally critical to global synchronization rather than relying on topology-dependent heuristics.

To examine whether FGIA’s advantage persists beyond purely topological effects, we extend the analysis to weighted networks. In this setting, each existing edge is assigned a random positive weight, and algebraic connectivity is computed from the corresponding weighted Laplacian. For each network configuration, all attack algorithms are repeated over 50 independent realizations of edge weights, and the resulting pRAC curves and AUC values are averaged. This protocol isolates the impact of interaction-strength heterogeneity while preserving the underlying network structure.

\begin{figure*}
    \centering
    \includegraphics[width=1\linewidth]{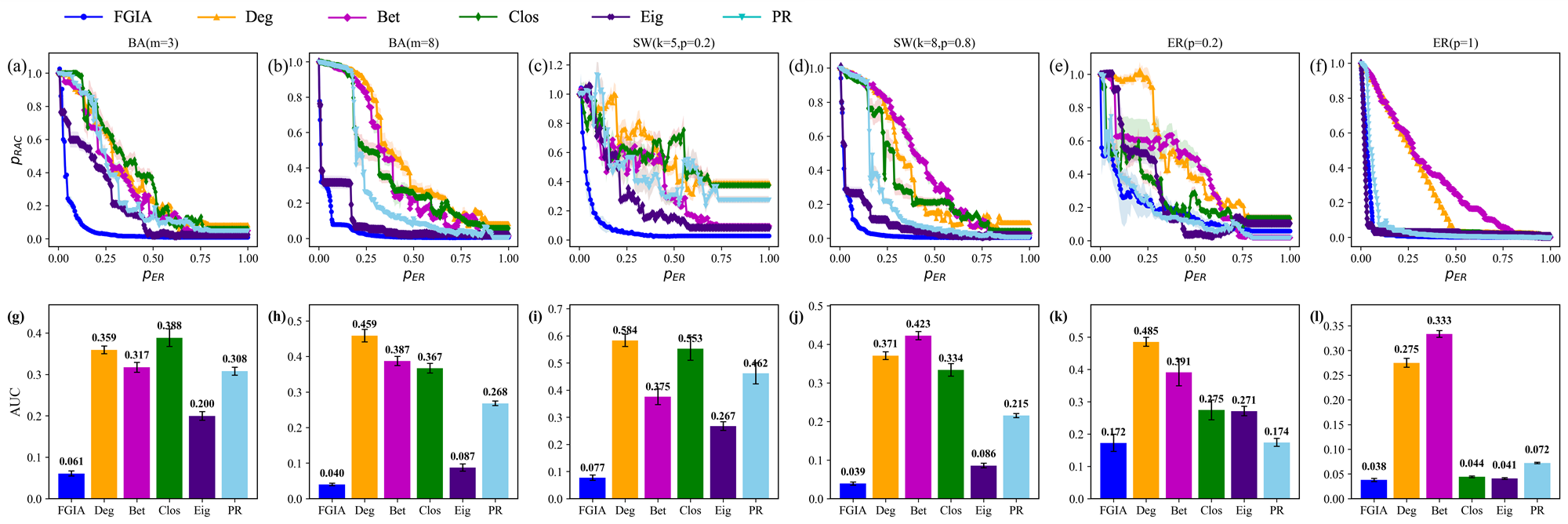}
    \caption{
Edge-removal performance on weighted synthetic networks.
Panels (a--f) show $p_{\mathrm{RAC}}$ versus $p_{\mathrm{ER}}$, and panels (g--l) report the corresponding AUC values.
Each directed edge is assigned an independent random weight
$w_{ij} \sim \mathrm{Uniform}(1-\varepsilon,\,1+\varepsilon)$ ($\varepsilon=0.3$),
which enters the robustness evaluation through the weighted directed Laplacian.
Results are averaged over 50 realizations; shaded regions denote one standard deviation.
}

    \label{fig:weighted_networks}
\end{figure*}

The results on weighted networks, summarized in Fig.~\ref{fig:weighted_networks}, show that FGIA remains consistently superior in this setting. Although weighted centrality measures partially adapt to heterogeneous edge strengths, their performance exhibits noticeable variability across realizations. In contrast, FGIA demonstrates both lower mean AUC values and reduced sensitivity to weight fluctuations. This robustness arises from the fact that FGIA exploits spectral gradients of the weighted Laplacian, enabling it to identify edges that are functionally indispensable to global synchronization even when interaction strengths are unevenly distributed.

\begin{figure*}
    \centering
    \includegraphics[width=1\linewidth]{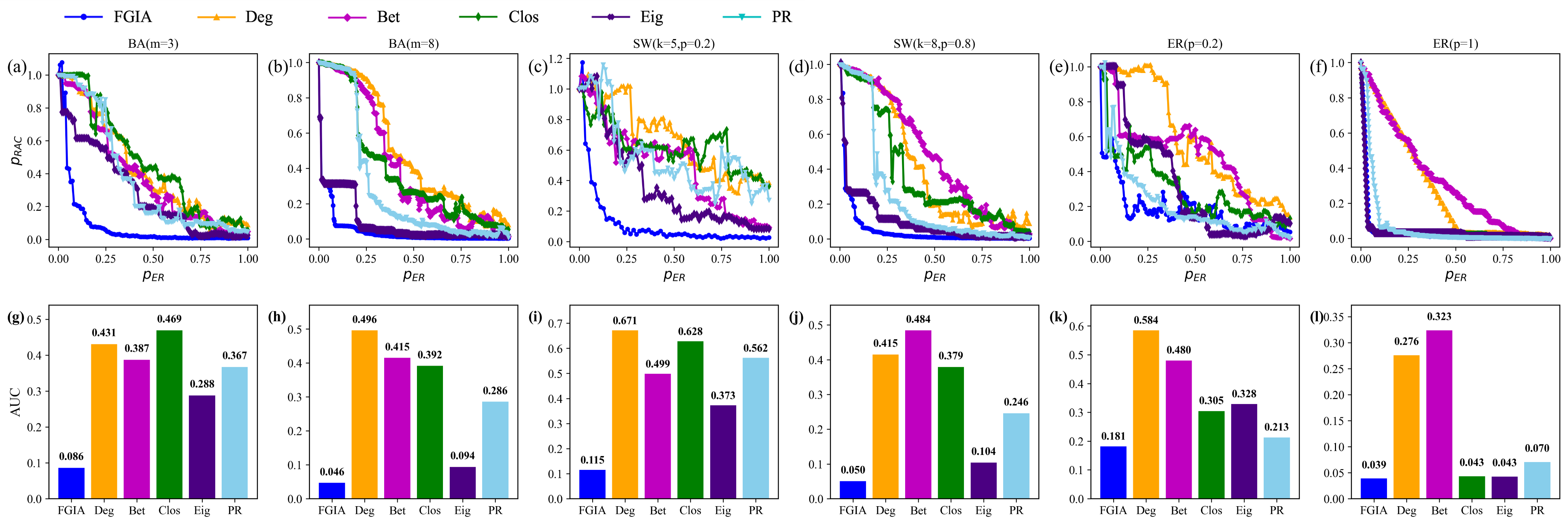}
    \caption{
    Edge-removal performance on directed synthetic networks.
    Panels (a--f) show pRAC as a function of pER for directed BA, WS, and ER networks.
    Panels (g--l) report the corresponding AUC values.
    }
    \label{fig:attack_algorithms_directed}
\end{figure*}

We further extend the analysis to directed networks, where edge directionality introduces asymmetric interactions and non-Hermitian spectral properties. Directed versions of BA, WS, and ER networks are constructed and constrained to remain strongly connected throughout the edge-removal process. Despite the increased structural complexity, the qualitative trends observed in undirected and weighted networks largely persist. FGIA continues to outperform all centrality-based baselines, achieving the most efficient degradation of algebraic connectivity across all tested directed network models, as shown in Fig.~\ref{fig:attack_algorithms_directed}.

A distinctive feature of directed networks is the possibility of counterintuitive spectral responses, whereby removing certain edges may temporarily increase algebraic connectivity. This phenomenon reflects the nontrivial organization of strongly connected components in directed graphs. To address this issue, FGIA incorporates an explicit filtering step that preserves global strong connectivity during edge removal. While this constraint can induce mild oscillations or transient increases in $\gamma$, the overall suppression trend remains dominant, and FGIA ultimately achieves the lowest AUC values among all compared methods.

Overall, the results on undirected, weighted, and directed networks consistently demonstrate the robustness and universality of FGIA. By leveraging gradient information derived from spectral properties rather than topology-specific centrality measures, FGIA reliably identifies edges whose removal most effectively suppresses global synchronization. Its effectiveness persists across network models, parameter regimes, weight heterogeneity, and edge directionality, establishing FGIA as a principled and efficient strategy for targeted disruption of complex networks.


\begin{figure*}
    \centering
    \includegraphics[width=1\linewidth]{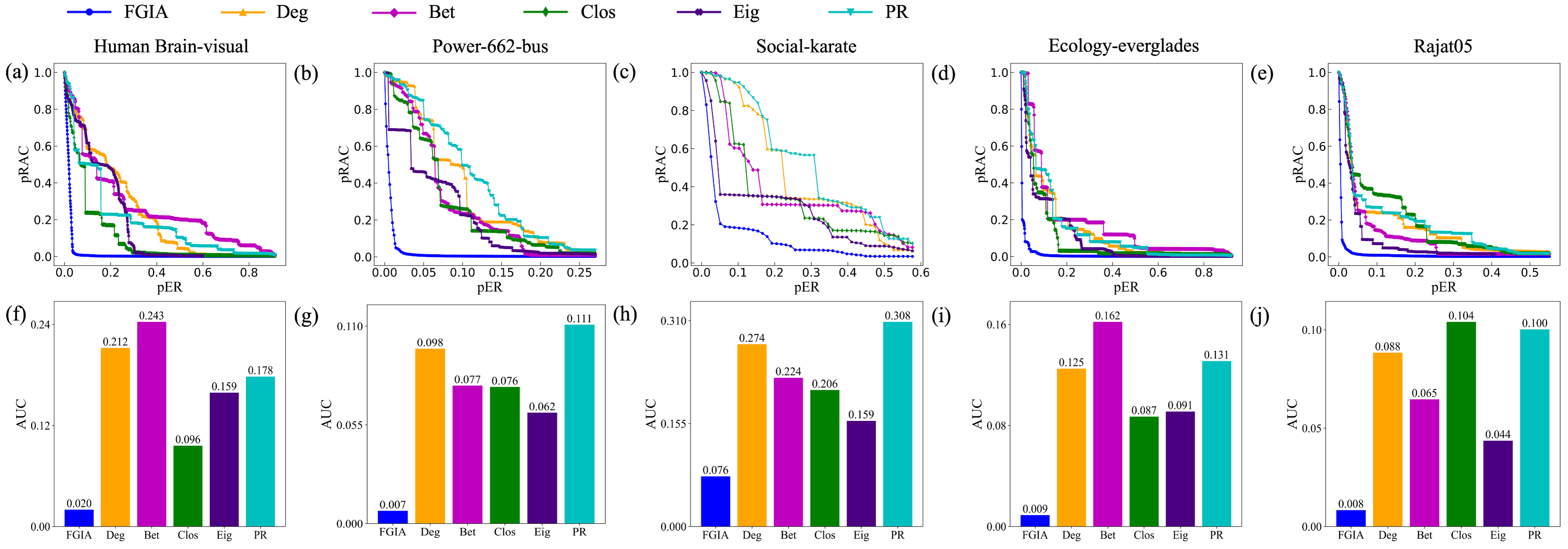}
    \caption{Edge-removal performance on real networks.
    Panels (a--f) show pRAC as a function of pER for directed BA, WS, and ER networks.
    Panels (g--l) report the corresponding AUC values.}
    \label{fig:real_app}
\end{figure*}

The edge-removal performance of different algorithms is next evaluated on real-world networks.
The results obtained on a representative subset of directed real networks are summarized in Fig.~\ref{fig:real_app},
covering systems from multiple application domains, including brain networks, power grids, social systems,
ecological networks, and other complex infrastructures.
The upper panels in Fig.~\ref{fig:real_app}(a--f) display the evolution of the normalized robustness indicator
$p_{\mathrm{RAC}}$ as a function of the removed edge fraction $p_{\mathrm{ER}}$,
while the lower panels in Fig.~\ref{fig:real_app}(g--l) report the corresponding AUC values,
which quantify the cumulative degradation of the synchronization indicator $\gamma$ during the attack process.

Unsurprisingly, the superiority of the proposed FGIA method is further validated across this broad range of real-world networks.
Across all representative examples shown in Fig.~\ref{fig:real_app},
FGIA consistently induces the fastest and strongest suppression of synchronization,
as evidenced by the steepest decay of $p_{\mathrm{RAC}}$ and the smallest AUC values.
This behavior persists despite substantial variations in network size, edge density, and functional context,
indicating that the effectiveness of FGIA is not tied to specific topological features.

To further substantiate this observation beyond the representative examples,
the same evaluation protocol is applied to a larger collection of real-world networks.
Table~\ref{tab:accuracy} summarizes the quantitative comparison,
reporting the AUC values obtained on a total of 26 datasets spanning six major categories
of complex systems, including brain networks, power systems, social networks,
interaction networks, ecological networks, and other real-world graphs.

Lower AUC values correspond to stronger overall suppression of synchronization.
For clarity, the smallest AUC value in each dataset is highlighted in red,
while the largest value is marked in blue.

As shown in Tab.~\ref{tab:accuracy},
FGIA achieves the lowest AUC in all datasets without exception,
consistently outperforming degree-, betweenness-, closeness-, eigenvector-,
and PageRank-based attack strategies.
This uniform dominance across heterogeneous real networks confirms the robustness
and scalability of FGIA in realistic settings.

As an illustrative example, in the visual sub-network of the human brain,
the quantitative results reported in Tab.~\ref{tab:accuracy} indicate that
FGIA achieves an AUC of $0.020$, which is $87.4\%$ lower than the second-best algorithm
(eigenvector centrality, AUC = $0.159$)
and $91.8\%$ lower than the worst-performing algorithm
(betweenness centrality, AUC = $0.243$).
Similar relative advantages can be observed across power grid and ecological networks,
further confirming the trends observed in Fig.~\ref{fig:real_app}
and the aggregate statistics in Tab.~\ref{tab:accuracy}.

At the same time, the efficacy of the compared algorithms varies significantly
depending on network-specific characteristics.
For example, betweenness centrality (Bet) exhibits relatively competitive performance
in the \textit{rajat05} network, as indicated in Fig.~\ref{fig:real_app}(e)
and corroborated by the corresponding AUC value in Tab.~\ref{tab:accuracy},
while it fails to generalize to the human visual brain network
in Fig.~\ref{fig:real_app}(a)
or the Everglades ecological network
in Fig.~\ref{fig:real_app}(d),
where its performance is among the weakest.
Such sensitivity reflects the inherent limitations of topology-dependent heuristics,
whose effectiveness relies strongly on particular structural patterns.

In contrast, all results consistently highlight the adaptability of FGIA.
By leveraging a gradient-driven spectral analysis,
FGIA bypasses the constraints of heuristic, topology-specific criteria
and directly targets global robustness degradation.
This mechanism ensures universal superiority regardless of network size, density,
or functional context, as consistently evidenced by both the representative examples
in Fig.~\ref{fig:real_app} and the comprehensive statistics in Tab.~\ref{tab:accuracy}.

\begin{table*}[!t]
    \centering
    \caption{AUC of different algorithms in real networks}
    \label{tab:accuracy}
        \begin{tabular}{lccrrrrrr}
        \toprule
         \textbf{Name} & Node Number $(n)$& Edge Number$(m)$& \textbf{FGIA} & Deg & Bet & Clos & Eig & PR \\
        \midrule
        \multicolumn{9}{l}{\textbf{Brain Network}} \\
        human-visual & 111 & 1276 & \color{red}\textbf{0.020} & 0.212 & \color{blue}0.243 & 0.096 & 0.159 & 0.178\\
        human-medial& 172 & 2542 & \color{red}\textbf{0.004} & 0.133 \color{blue}& \color{blue}0.185& 0.057 & 0.154 & 0.131\\
        macaque-rhesus-cerebral-cortex 1&91&1401&\color{red}\textbf{0.009}&0.111&\color{blue}0.139&0.106&0.077&0.123\\
        macaque-rhesus-brain 2&91&582&\color{red}\textbf{0.152}&0.609&0.613&0.491&0.502&\color{blue}0.626\\
        mouse-visual-cortex 2&193&214&\color{red}\textbf{0.024}&\color{blue}0.071&0.022&0.033&0.052&0.044\\
        cat-mixed-species-brain 1&65&730&\color{red}\textbf{0.015}&0.132&\color{blue}0.199&0.075&0.081&0.144\\

        \midrule
        \multicolumn{9}{l}{\textbf{Power Networks}} \\
        power-494-bus & 494 & 586 & \color{red}\textbf{0.010} & \color{blue}0.115& 0.018 & 0.080 & 0.059 & 0.091\\
        power-662-bus & 662 & 906 & \color{red}\textbf{0.007} & 0.098& 0.077& 0.076& 0.062& \color{blue}0.111\\
        power-bcspwr09 & 1723 & 4117& \color{red}\textbf{0.005} & 0.070& 0.021& 0.052& 0.042& \color{blue}0.076\\

        \midrule
        \multicolumn{9}{l}{\textbf{Social Networks}} \\
        soc-tribes&16&58&\color{red}\textbf{0.074}&0.171&0.294&0.369&0.447&\color{blue}0.458\\
        soc-karate&34 & 78 & \color{red}\textbf{0.076} &0.274 &0.224&0.206&0.159&\color{blue}0.308 \\
        soc-dolphins&62 & 159 & \color{red}\textbf{0.039} & 0.239& 0.040& 0.216& 0.192& \color{blue}0.318\\
        soc-pages-food&620 & 2102 & \color{red}\textbf{0.018} & 0.310& 0.019& 0.232& 0.298& \color{blue}0.311\\

        \midrule
        \multicolumn{9}{l}{\textbf{Interaction Networks}} \\
        enron-only&143 & 623 & \color{red}\textbf{0.016} & 0.348&0.041 &0.204 &0.100 &\color{blue}0.428 \\
        crime-moreno&829 & 1476 & \color{red}\textbf{0.002} & \color{blue}0.156& 0.149& 0.060& 0.113&\color{blue}0.156 \\

        \midrule
         \multicolumn{9}{l}{\textbf{Ecology Networks}} \\
         eco-stmarks&54&350&\color{red}\textbf{0.016}&0.144&\color{blue}0.176&0.099&0.139&0.146\\
         eco-everglades&69&880&\color{red}\textbf{0.009}&0.125&\color{blue}0.162&0.087&0.091&0.131\\
         eco-mangwet&97&1446&\color{red}\textbf{0.008}&0.088&0.094&0.040&\color{blue}0.099&0.074\\
         eco-florida&128&2075&\color{red}\textbf{0.007}&0.224&\color{blue}0.268&0.073&0.262&0.254\\
         eco-foodweb-baydry&128&2106&\color{red}\textbf{0.007}&\color{blue}0.293&0.283&0.067&0.266&0.234\\

        \midrule
         \multicolumn{9}{l}{\textbf{Other Networks}} \\
        road-Chesapeake&39&170&\color{red}\textbf{0.026}&0.168&\color{blue}0.282&0.120&0.124&0.202\\
        rajat05&301&668&\color{red}\textbf{0.008}&0.088&0.065&\color{blue}0.104&0.044&0.100\\
        insecta-ant-colony2-day37&58&928&\color{red}\textbf{0.009}&0.028&0.177&0.648&0.658&\color{blue}0.693\\
         ca-netscience&379&914&\color{red}\textbf{0.016}&\color{blue}0.235&0.021&0.019&0.164&0.106\\
         bio-celegansneural&297&2148&\color{red}\textbf{0.011}&0.270&0.064&0.300&\color{blue}0.452&0.112\\
        infect-hyper&113 & 2196 & \color{red}\textbf{0.050} & \color{blue}0.483& 0.143& 0.193&0.402 &0.357 \\

        \bottomrule
    \end{tabular}
\end{table*}

\section{Conclusion}

The Fiedler Gradient Iterative Attack (FGIA) algorithm is proposed as a systematic framework for suppressing synchronization in complex networks, where synchronization performance is quantified by the generalized Fiedler value $\gamma$ of the Laplacian matrix. Since a larger $\gamma$ corresponds to faster convergence toward the synchronized state and stronger collective coherence, reducing $\gamma$ effectively weakens synchronization robustness and delays dynamical recovery. By leveraging spectral perturbation theory, FGIA identifies edges whose removal induces the largest degradation of $\gamma$, using the Fiedler vector gradient $\nabla \epsilon_{ij}$ as a physically interpretable measure of edge-level contribution to global synchronization.

The proposed algorithm integrates hierarchical spectral bisection, gradient-based edge ranking, and bridge detection to iteratively remove synchronization-critical edges while explicitly preserving global connectivity. Compared with brute-force optimization strategies, FGIA achieves substantially lower computational complexity, enabling scalable application to large-scale networks without sacrificing effectiveness. Extensive numerical results on synthetic benchmarks demonstrate that FGIA consistently outperforms conventional centrality-based attack strategies across a wide range of topological regimes.

Importantly, the analysis presented in this work encompasses all classical categories of static networks, including random (Erd\H{o}s--R\'enyi), small-world (Watts--Strogatz), and scale-free (Barab\'asi--Albert) models, as well as their directed and weighted extensions. The effectiveness of FGIA is further validated on diverse real-world networks drawn from multiple application domains, such as brain connectomes, power grids, ecological food webs, and social interaction networks. These results indicate that the proposed spectral-gradient framework provides a robust and topology-agnostic mechanism for identifying synchronization vulnerabilities in realistic networked systems.

This work bridges fundamental concepts from spectral graph theory with algorithmic strategies for network control, demonstrating that synchronization robustness in a broad class of static networks can be systematically weakened through localized and spectrally informed edge removals. While the present formulation focuses on strongly synchronized regimes in which linear spectral metrics remain predictive of collective dynamics, it establishes a clear and physically grounded connection between network structure and synchronization stability.

Viewed within the broader literature on synchronization stability, robustness and stability of synchronization in complex networks with higher-dimensional node dynamics and more intricate collective behaviors have been extensively investigated in previous works~\cite{Barahona2002,Nishikawa2003PRL,pecora1998master}. In particular, the Master Stability Function (MSF) framework provides a general theoretical basis for analyzing synchronization stability in networks of identical oscillators with multi-dimensional dynamics~\cite{pecora1998master}, while symmetry-based approaches have revealed the emergence of remote synchronization, cluster synchronization, partial synchronization, and chimera states in structured networks~\cite{bergner2012remote,Nicosia2013Remote,luo2024effects,luo2024simplex,abrams2004chimera,sethia2014existence,tinsley2012chimera,pecora2014cluster,LUO2026117581,golubitsky2016rigid,martens2013spiral}. From this perspective, the present work, which characterizes synchronization robustness through the generalized Fiedler value of the Laplacian, addresses a specific but fundamental regime in which collective stability is governed by the slowest mode near the synchronized manifold~\cite{menck2013basin}. Within this regime, the proposed spectral-gradient framework offers a structurally grounded and computationally efficient approach for assessing and suppressing synchronization robustness, and its conclusions are fully consistent with more general stability theories when restricted to strongly synchronized, diffusively coupled systems.

Taken together, these results establish FGIA as a principled and efficient tool for analyzing and controlling synchronization robustness in complex networks. By explicitly linking edge-level structural perturbations to global spectral stability, the proposed approach provides new insight into how localized modifications can induce macroscopic dynamical consequences, offering a unified perspective on synchronization vulnerability across a wide range of networked systems.

\section*{Data Availability}
All real-world network datasets employed in this study are publicly available from the Network Data Repository~\cite{networkdata}. The datasets encompass diverse domains, including brain functional networks, power grid systems, social interaction graphs, ecological food webs, and infrastructure networks, as summarized in Tab.~\ref{tab:accuracy}. No proprietary or restricted data were used, and no additional preprocessing beyond standard symmetrization and normalization was applied. Consequently, the results presented here are fully reproducible using the openly accessible datasets.

% \section{Acknowledgment}

% No acknowledgment.

\bibliography{Ref_Luo}

@article{Arenas2008,
  title = {Synchronization in complex networks},
  author = {Arenas, Alex and D{\'i}az-Guilera, Albert and Kurths, J{\"u}rgen and Moreno, Yamir and Zhou, Changsong},
  journal = {Physics Reports},
  volume = {469},
  number = {3},
  pages = {93--153},
  year = {2008},
  doi = {10.1016/j.physrep.2008.09.002}
}

@article{Boccaletti2006PhysRep,
  title = {Complex networks: Structure and dynamics},
  author = {Boccaletti, Stefano and Latora, Vito and Moreno, Yamir and Chavez, Martin and Hwang, D.-U.},
  journal = {Physics Reports},
  volume = {424},
  number = {4--5},
  pages = {175--308},
  year = {2006},
  doi = {10.1016/j.physrep.2005.10.009}
}

@article{Barahona2002,
  title = {Synchronization in small-world systems},
  author = {Barahona, Mauricio and Pecora, Louis M.},
  journal = {Physical Review Letters},
  volume = {89},
  pages = {054101},
  year = {2002},
  doi = {10.1103/PhysRevLett.89.054101}
}

@article{Qi2024NetworkResilience,
  title = {Network Resilience: Definitions, approaches, and applications},
  author = {Qi, Xiaoyu and Mei, Gang},
  journal = {Journal of King Saud University - Computer and Information Sciences},
  volume = {36},
  number = {1},
  pages = {101882},
  year = {2024},
  doi = {10.1016/j.jksuci.2023.101882}
}

@article{Motter2005PRE,
  title = {Network synchronization, diffusion, and the paradox of heterogeneity},
  author = {Motter, Adilson E. and Zhou, Changsong and Kurths, J{\"u}rgen},
  journal = {Physical Review E},
  volume = {71},
  pages = {016116},
  year = {2005},
  doi = {10.1103/PhysRevE.71.016116}
}

@article{Motter2015,
  title = {Networkcontrology},
  author = {Motter, Adilson E.},
  journal = {Chaos: An Interdisciplinary Journal of Nonlinear Science},
  volume = {25},
  number = {9},
  pages = {097621},
  year = {2015},
  doi = {10.1063/1.4931570}
}

@article{Belykh2005synchronization,
  title = {Synchronization of bursting neurons: What matters in the network topology},
  author = {Belykh, Igor and de Lange, Enno and Hasler, Martin},
  journal = {Physical Review Letters},
  volume = {94},
  pages = {188101},
  year = {2005},
  doi = {10.1103/PhysRevLett.94.188101}
}

@article{Galinsky2023critically,
  title = {Critically synchronized brain waves form an effective, robust and flexible basis for human memory and learning},
  author = {Galinsky, Vitaly L. and Frank, Lawrence R.},
  journal = {Scientific Reports},
  volume = {13},
  pages = {4343},
  year = {2023},
  doi = {10.1038/s41598-023-31365-6}
}

@article{mathiopoulou2025gamma,
  title = {Gamma entrainment induced by deep brain stimulation as a biomarker for motor improvement with neuromodulation},
  author = {Mathiopoulou, Varvara and Habets, Jeroen and Feldmann, Lucia K. and Busch, Johannes L. and Roediger, Jan and Behnke, Jennifer K. and Schneider, Gerd-Helge and Faust, Katharina and K{\"u}hn, Andrea A.},
  journal = {Nature Communications},
  volume = {16},
  number = {1},
  pages = {2956},
  year = {2025},
  doi = {10.1038/s41467-025-58132-7}
}

@article{ruan2023super,
  title = {Super-Resolution Perception Assisted Spatiotemporal Graph Deep Learning Against False Data Injection Attacks in Smart Grid},
  author = {Ruan, Jiaqi and Fan, Gang and Zhu, Yifan and Liang, Gaoqi and Zhao, Junhua and Wen, Fushuan and Dong, Zhao Yang},
  journal = {IEEE Transactions on Smart Grid},
  volume = {14},
  number = {5},
  pages = {4035--4046},
  year = {2023},
  doi = {10.1109/TSG.2023.3241268}
}

@article{mishra2023resilience,
  title = {Resilience-Driven Scheme in Multiple Microgrids With Secure Transactive Energy System Framework},
  author = {Mishra, Dillip Kumar and Wang, Jiatong and Li, Li and Zhang, Jiangfeng and Hossain, M. J.},
  journal = {IEEE Transactions on Industry Applications},
  volume = {60},
  number = {2},
  pages = {2277--2289},
  year = {2024},
  doi = {10.1109/TIA.2023.3299260}
}

@incollection{Tsang2019,
  title = {Enhancing Synchronization Stability in Complex Networks with Probabilistic Natural Frequencies},
  author = {Tsang, K. Y. Henry and Li, Bo and Wong, K. Y. Michael},
  booktitle = {Complex Networks and Their Applications VII},
  editor = {Aiello, Luca Maria and Cherifi, Chantal and Cherifi, Hocine and Lambiotte, Renaud and Li{\'o}, Pietro and Rocha, Luis M.},
  series = {Studies in Computational Intelligence},
  publisher = {Springer International Publishing},
  address = {Cham},
  pages = {854--865},
  year = {2019},
  doi = {10.1007/978-3-030-05411-3_68}
}

@article{Richard202Synchronization,
  title = {A Fast Edge-Based Synchronizer for Tasks in Real-Time Artificial Intelligence Applications},
  author = {Olaniyan, Richard and Maheswaran, Muthucumaru},
  journal = {IEEE Internet of Things Journal},
  volume = {9},
  number = {5},
  pages = {3825--3837},
  year = {2022},
  doi = {10.1109/JIOT.2021.3100295}
}

@article{chen2015saturated,
  title = {Saturated {Nussbaum} Function Based Approach for Robotic Systems With Unknown Actuator Dynamics},
  author = {Chen, Ci and Liu, Zhi and Zhang, Yun and Chen, C. L. Philip and Xie, Shengli},
  journal = {IEEE Transactions on Cybernetics},
  volume = {46},
  number = {10},
  pages = {2311--2322},
  year = {2016},
  doi = {10.1109/TCYB.2015.2475363}
}

@inproceedings{pan2020justinian,
  title = {Justinian's {GAAvernor}: Robust Distributed Learning with Gradient Aggregation Agent},
  author = {Pan, Xudong and Zhang, Mi and Wu, Duocai and Xiao, Qifan and Ji, Shouling and Yang, Zhemin},
  booktitle = {29th USENIX Security Symposium (USENIX Security 20)},
  publisher = {USENIX Association},
  isbn = {978-1-939133-17-5},
  pages = {1641--1658},
  year = {2020},
  url = {https://www.usenix.org/conference/usenixsecurity20/presentation/pan}
}

@article{Arenas2006,
  title = {Synchronization reveals topological scales in complex networks},
  author = {Arenas, Alex and D{\'i}az-Guilera, Albert and P{\'e}rez-Vicente, Conrad J.},
  journal = {Physical Review Letters},
  volume = {96},
  pages = {114102},
  year = {2006},
  doi = {10.1103/PhysRevLett.96.114102}
}

@article{Crucitti2004,
  title = {Error and attack tolerance of complex networks},
  author = {Crucitti, Paolo and Latora, Vito and Marchiori, Massimo and Rapisarda, Andrea},
  journal = {Physica A: Statistical Mechanics and its Applications},
  volume = {340},
  number = {1--3},
  pages = {388--394},
  year = {2004},
  doi = {10.1016/j.physa.2004.04.031}
}

@article{Albert2000Attack,
  title = {Error and attack tolerance of complex networks},
  author = {Albert, R{\'e}ka and Jeong, Hawoong and Barab{\'a}si, Albert-L{\'a}szl{\'o}},
  journal = {Nature},
  volume = {406},
  pages = {378--382},
  year = {2000},
  doi = {10.1038/35019019}
}

@article{fiedler1973algebraic,
  title = {Algebraic connectivity of graphs},
  author = {Fiedler, Miroslav},
  journal = {Czechoslovak Mathematical Journal},
  volume = {23},
  number = {2},
  pages = {298--305},
  year = {1973},
  doi = {10.21136/CMJ.1973.101168}
}

@article{Dorfler2013Sync,
  title = {Synchronization in complex oscillator networks and smart grids},
  author = {D{\"o}rfler, Florian and Chertkov, Michael and Bullo, Francesco},
  journal = {Proceedings of the National Academy of Sciences of the United States of America},
  volume = {110},
  number = {6},
  pages = {2005--2010},
  year = {2013},
  doi = {10.1073/pnas.1212134110}
}

@article{chen2024node,
  title = {Node Centrality Based on Edge Dynamics in a Chaotic Network},
  author = {Chen, Fangyue and Zhou, Jin and Lu, Jun-an},
  journal = {International Journal of Bifurcation and Chaos},
  volume = {34},
  number = {11},
  pages = {2450142},
  year = {2024},
  doi = {10.1142/S0218127424501426}
}

@article{jiang2024fiedler,
  title = {Fiedler value: The cumulated dynamical contribution value of all edges in a complex network},
  author = {Jiang, Siyang and Lu, Jun-an and Zhou, Jin and Dai, Qinrui},
  journal = {Physical Review E},
  volume = {109},
  pages = {054301},
  year = {2024},
  doi = {10.1103/PhysRevE.109.054301}
}

@article{jiang2023searching,
  title = {Searching for Key Cycles in a Complex Network},
  author = {Jiang, Siyang and Zhou, Jin and Small, Michael and Lu, Jun-an and Zhang, Yanqi},
  journal = {Physical Review Letters},
  volume = {130},
  pages = {187402},
  year = {2023},
  doi = {10.1103/PhysRevLett.130.187402}
}

@article{ZhangZeng2014PhysicaA,
  title = {Network skeleton for synchronization: Identifying redundant connections},
  author = {Zhang, Cheng-Jun and Zeng, An},
  journal = {Physica A: Statistical Mechanics and its Applications},
  volume = {402},
  pages = {180--185},
  year = {2014},
  doi = {10.1016/j.physa.2014.02.002}
}

@article{Richter2023SpectralDynamics,
  title = {Spectral dynamics of guided edge removals and identifying transient amplifiers for death--Birth updating},
  author = {Richter, Hendrik},
  journal = {Journal of Mathematical Biology},
  volume = {87},
  number = {1},
  pages = {3},
  year = {2023},
  doi = {10.1007/s00285-023-01937-1}
}

@article{MilaneseSunNishikawa2010PRE,
  title = {Approximating spectral impact of structural perturbations in large networks},
  author = {Milanese, Attilio and Sun, Jie and Nishikawa, Takashi},
  journal = {Physical Review E},
  volume = {81},
  pages = {046112},
  year = {2010},
  doi = {10.1103/PhysRevE.81.046112}
}

@incollection{ChenLuZhanChen2012Chapter,
  title = {Laplacian Spectra and Synchronization Processes on Complex Networks},
  author = {Chen, Juan and Lu, Jun-an and Zhan, Choujun and Chen, Guanrong},
  booktitle = {Handbook of Optimization in Complex Networks},
  editor = {Thai, My T. and Pardalos, Panos M.},
  series = {Springer Optimization and Its Applications},
  publisher = {Springer},
  pages = {81--113},
  year = {2012},
  doi = {10.1007/978-1-4614-0754-6_4}
}

@article{YangJiang2017PhysicaA,
  title = {Impacts of link addition and removal on synchronization of an elementary power network},
  author = {Yang, Li-xin and Jiang, Jun},
  journal = {Physica A: Statistical Mechanics and its Applications},
  volume = {479},
  pages = {99--107},
  year = {2017},
  doi = {10.1016/j.physa.2017.02.055}
}

@article{ZengLuZhou2012NJP,
  title = {Manipulating directed networks for better synchronization},
  author = {Zeng, An and L{\"u}, Linyuan and Zhou, Tao},
  journal = {New Journal of Physics},
  volume = {14},
  pages = {083006},
  year = {2012},
  doi = {10.1088/1367-2630/14/8/083006}
}

@book{Chung1997SpectralGraphTheory,
  title = {Spectral Graph Theory},
  author = {Chung, Fan R. K.},
  publisher = {American Mathematical Society},
  address = {Providence, RI},
  series = {CBMS Regional Conference Series in Mathematics},
  volume = {92},
  year = {1997}
}

@incollection{Mohar1991LaplacianSpectrum,
  title = {The Laplacian Spectrum of Graphs},
  author = {Mohar, Bojan},
  booktitle = {Graph Theory, Combinatorics, and Applications},
  editor = {Alavi, Y. and Chartrand, G. and Oellermann, O. R. and Schwenk, A. J.},
  publisher = {Wiley},
  address = {New York},
  volume = {2},
  pages = {871--898},
  year = {1991}
}

@incollection{Kuramoto1975,
  title = {Self-entrainment of a population of coupled non-linear oscillators},
  author = {Kuramoto, Yoshiki},
  booktitle = {International Symposium on Mathematical Problems in Theoretical Physics},
  series = {Lecture Notes in Physics},
  volume = {39},
  publisher = {Springer},
  pages = {420--422},
  year = {1975},
  doi = {10.1007/BFb0013365}
}

@article{olfati2004consensus,
  title = {Consensus problems in networks of agents with switching topology and time-delays},
  author = {Olfati-Saber, Reza and Murray, Richard M.},
  journal = {IEEE Transactions on Automatic Control},
  volume = {49},
  number = {9},
  pages = {1520--1533},
  year = {2004},
  doi = {10.1109/TAC.2004.834113}
}

@article{jadbabaie2003coordination,
  title = {Coordination of groups of mobile autonomous agents using nearest neighbor rules},
  author = {Jadbabaie, Ali and Lin, Jie and Morse, A. Stephen},
  journal = {IEEE Transactions on Automatic Control},
  volume = {48},
  number = {6},
  pages = {988--1001},
  year = {2003},
  doi = {10.1109/TAC.2003.812781}
}

@article{li2010consensus,
  title = {Consensus conditions of multi-agent systems with time-varying topologies and stochastic communication noises},
  author = {Li, Tao and Zhang, Ji-Feng},
  journal = {IEEE Transactions on Automatic Control},
  volume = {55},
  number = {9},
  pages = {2043--2057},
  year = {2010},
  doi = {10.1109/TAC.2010.2042982}
}

@article{Wu2005,
  title = {Algebraic connectivity of directed graphs},
  author = {Wu, Chai Wah},
  journal = {Linear and Multilinear Algebra},
  volume = {53},
  number = {3},
  pages = {203--223},
  year = {2005},
  doi = {10.1080/03081080500054810}
}

@book{Wu2007,
  title = {Synchronization in Complex Networks of Nonlinear Dynamical Systems},
  author = {Wu, Chai Wah},
  publisher = {World Scientific},
  address = {Singapore},
  year = {2007}
}

@article{ImaeCai2018,
  title = {On algebraic connectivity of directed scale-free networks},
  author = {Imae, Takanobu and Cai, Kai},
  journal = {Journal of the Franklin Institute},
  volume = {355},
  number = {16},
  pages = {8065--8078},
  year = {2018},
  doi = {10.1016/j.jfranklin.2018.07.038}
}

@article{liu2022network,
  title = {Network resilience},
  author = {Liu, Xueming and Li, Daqing and Ma, Manqing and Szymanski, Boleslaw K. and Stanley, H. Eugene and Gao, Jianxi},
  journal = {Physics Reports},
  volume = {971},
  pages = {1--108},
  year = {2022},
  doi = {10.1016/j.physrep.2022.04.002}
}

@article{Gao2016Universal,
  title = {Universal resilience patterns in complex networks},
  author = {Gao, Jianxi and Barzel, Baruch and Barab{\'a}si, Albert-L{\'a}szl{\'o}},
  journal = {Nature},
  volume = {530},
  pages = {307--312},
  year = {2016},
  doi = {10.1038/nature16948}
}

@article{Tarjan72Depth,
  title = {Depth-first search and linear graph algorithms},
  author = {Tarjan, Robert Endre},
  journal = {SIAM Journal on Computing},
  volume = {1},
  number = {2},
  pages = {146--160},
  year = {1972},
  doi = {10.1137/0201010}
}

@article{Nishikawa2003PRL,
  title = {Heterogeneity in Oscillator Networks: Are Smaller Worlds Easier to Synchronize?},
  author = {Nishikawa, Takashi and Motter, Adilson E. and Lai, Ying-Cheng and Hoppensteadt, Frank C.},
  journal = {Physical Review Letters},
  volume = {91},
  pages = {014101},
  year = {2003},
  doi = {10.1103/PhysRevLett.91.014101}
}

@article{pecora1998master,
  title = {Master Stability Functions for Synchronized Coupled Systems},
  author = {Pecora, Louis M. and Carroll, Thomas L.},
  journal = {Physical Review Letters},
  volume = {80},
  number = {10},
  pages = {2109--2112},
  year = {1998},
  doi = {10.1103/PhysRevLett.80.2109}
}

@article{bergner2012remote,
  title = {Remote synchronization in star networks},
  author = {Bergner, A. and Frasca, M. and Sciuto, G. and Buscarino, A. and Ngamga, E. J. and Fortuna, L. and Kurths, J.},
  journal = {Physical Review E},
  volume = {85},
  pages = {026208},
  year = {2012},
  doi = {10.1103/PhysRevE.85.026208}
}

@article{Nicosia2013Remote,
  title = {Remote Synchronization Reveals Network Symmetries and Functional Modules},
  author = {Nicosia, Vincenzo and Valencia, Miguel and Chavez, Mario and D{\'i}az-Guilera, Albert and Latora, Vito},
  journal = {Physical Review Letters},
  volume = {110},
  pages = {174102},
  year = {2013},
  doi = {10.1103/PhysRevLett.110.174102}
}

@article{luo2024effects,
  title = {Effects of uncommon non-isochronicities on remote synchronization},
  author = {Luo, Kaiming and Cai, Zongkai and Liu, Zonghua and Guan, Shuguang and Zou, Yong},
  journal = {Chaos, Solitons \& Fractals},
  volume = {181},
  pages = {114705},
  year = {2024},
  doi = {10.1016/j.chaos.2024.114705}
}

@article{luo2024simplex,
  title = {Reconstruction of simplex structures based on phase synchronization dynamics},
  author = {Luo, Kai-Ming and Guan, Shu-Guang and Zou, Yong},
  journal = {Acta Physica Sinica},
  volume = {73},
  number = {12},
  pages = {120501},
  year = {2024},
  doi = {10.7498/aps.73.20240334}
}

@article{abrams2004chimera,
  title = {Chimera states for coupled oscillators},
  author = {Abrams, Daniel M. and Strogatz, Steven H.},
  journal = {Physical Review Letters},
  volume = {93},
  pages = {174102},
  year = {2004},
  doi = {10.1103/PhysRevLett.93.174102}
}

@article{sethia2014existence,
  title = {Chimera states: The existence criteria revisited},
  author = {Sethia, Gautam C. and Sen, Abhijit},
  journal = {Physical Review Letters},
  volume = {112},
  pages = {144101},
  year = {2014},
  doi = {10.1103/PhysRevLett.112.144101}
}

@article{tinsley2012chimera,
  title = {Chimera and phase-cluster states in populations of coupled chemical oscillators},
  author = {Tinsley, Mark R. and Nkomo, Simbarashe and Showalter, Kenneth},
  journal = {Nature Physics},
  volume = {8},
  pages = {662--665},
  year = {2012},
  doi = {10.1038/nphys2371}
}

@article{pecora2014cluster,
  title = {Cluster synchronization and isolated desynchronization in complex networks with symmetries},
  author = {Pecora, Louis M. and Sorrentino, Francesco and Hagerstrom, Aaron M. and Murphy, Thomas E. and Roy, Rajarshi},
  journal = {Nature Communications},
  volume = {5},
  pages = {4079},
  year = {2014},
  doi = {10.1038/ncomms5079}
}

@article{LUO2026117581,
  title = {Hierarchical synchronization and distortion scaling in social media networks: A fractal-like topology theory},
  author = {Luo, Kaiming},
  journal = {Chaos, Solitons \& Fractals},
  volume = {202},
  pages = {117581},
  year = {2026},
  doi = {10.1016/j.chaos.2025.117581}
}

@article{golubitsky2016rigid,
  title = {Rigid patterns of synchrony for equilibria and periodic cycles in network dynamics},
  author = {Golubitsky, Martin and Stewart, Ian},
  journal = {Chaos: An Interdisciplinary Journal of Nonlinear Science},
  volume = {26},
  pages = {094803},
  year = {2016},
  doi = {10.1063/1.4953664}
}

@article{martens2013spiral,
  title = {Solvable model of spiral-wave chimeras},
  author = {Martens, Erik A. and Laing, Carlo R. and Strogatz, Steven H.},
  journal = {Physical Review Letters},
  volume = {104},
  pages = {044101},
  year = {2010},
  doi = {10.1103/PhysRevLett.104.044101}
}

@article{menck2013basin,
  title = {How basin stability complements the linear-stability paradigm},
  author = {Menck, Peter J. and Heitzig, Jobst and Marwan, Norbert and Kurths, J{\"u}rgen},
  journal = {Nature Physics},
  volume = {9},
  pages = {89--92},
  year = {2013},
  doi = {10.1038/nphys2516}
}

@inproceedings{networkdata,
  title = {The Network Data Repository with Interactive Graph Analytics and Visualization},
  author = {Rossi, Ryan A. and Ahmed, Nesreen K.},
  booktitle = {Proceedings of the AAAI Conference on Artificial Intelligence},
  volume = {29},
  number = {1},
  pages = {4292--4293},
  year = {2015},
  doi = {10.1609/aaai.v29i1.9277},
  url = {https://networkrepository.com}
}

\appendix
\section{Benchmark Attack Algorithms}
\label{app:algorithms}

Here, we implement five compared attack algorithms, with their key procedures formalized briefly in Alg.~\ref{alg:compared}.

This algorithm aims to identify the set of edges to attack in a given network graph by utilizing node benchmarks. Initially, node benchmarks are computed based on various centrality measures, including Degree (Deg), Betweenness centrality(Bet), Closeness centrality(Clos), Eigenvector centrality(Eig), and PageRank (PR). 

\begin{algorithm}
\caption{Benchmark-based Attack}
\label{alg:compared}
\SetKwInOut{Input}{Input}
\SetKwInOut{Output}{Output}
\Input{Graph $G=(V,E)$, attack budget $k$}
\Output{Edge set $E_{\text{attack}}$}
\BlankLine
Compute node benchmarks (Deg, Bet, Clos, Eig, PR) for all $v_i \in V$\\
Initialize edge sets $E_{\text{sum}}$, $E_{\text{diff}}$ and corresponding graphs\\
\For{$\mathrm{strategy} \in \{\textnormal{sum},\textnormal{diff}\}$}{
    \For{$t \leftarrow 1$ \KwTo $k$}{
        Partition $G_{\mathrm{strategy}}$ into two communities via Fiedler vector\\
        Identify cross-community edges and exclude bridges (Tarjan's algorithm)\\
        Rank remaining candidate edges using
        \[
        \phi_{uv} =
        \begin{cases}
        B_u + B_v, & \text{sum strategy},\\
        |B_u - B_v|, & \text{diff strategy},
        \end{cases}
        \]
        Select highest-ranked edge preserving connectivity and remove it\\
    }
}
Evaluate AUC for both strategies and return the better result\\
\end{algorithm}

Here, two approaches are considered: \textit{sum} and \textit{diff}  in order to ensure a fair comparison with the optimal attack effects achievable by each centrality-based indicator. . Similar with Alg.~\ref{alg:FGIA}, for each strategy, the graph is partitioned using the Fiedler vector, and cross-edges between the partitions are identified. Bridge edges are detected using Tarjan’s algorithm and filtered out from the set of candidate edges. The remaining candidate edges are sorted based on the sum or the absolute difference of the node benchmarks, depending on the AUC performance. each candidate edge is then tested by temporarily removing it from the graph and checking if the graph remains connected. The first valid edge that satisfies this condition is selected. After executing this process for $k$ iterations, the algorithm evaluates the attack results for both strategies using the AUC metric. Finally, the algorithm returns the edge set that minimizes the AUC value, which represents the optimal set of edges for the attack. This approach exploits the network's structural vulnerabilities by combining node benchmarks with partitioning strategies to identify critical edges for removal. 

Therefore, this multi-strategy algorithmic approach guarantees the completeness of edge metrics constructed from classical node benchmarks, thereby establishing a more rigorous comparative basis for benchmarking against the Fiedler Gradient Iterative Attack (FGIA).

\end{document}